# Measuring the Impact of Missingness in Traffic Stop Data

**Saatvik Kher**[1] **Johanna Hardin**[2]

[1]Department of Computer Science, University of California Irvine, Irvine, California, United States of America,
[2]Mathematics and Statistics Department, Pomona College, Claremont, California, United States of America

**ABSTRACT**

In this article we explore the data available through the Stanford Open Policing Project. The data consist of information on millions of traffic stops across close to 100 different cities and highway patrols. Using a variety of metrics, we identify that the data is not missing completely at random. Furthermore, we develop ways of quantifying and visualizing missingness trends for different variables across the data sets. Because of the way we have identified the missingness dependence on key variables, imputation is not possible and a sensitivity analysis is required. We perform a sensitivity analysis to extend work done on the outcome test as well as to extend work done on sharp bounds on the average treatment effect. We demonstrate that bias calculations can fundamentally shift depending on the assumptions made about the observations for which the race variable has not been recorded. We suggest ways that our missingness sensitivity analysis can be extended to myriad contexts.

---

## Media Summary

Traffic stop data is collected nationally, partly motivated as a way to assess and address issues of racial profiling. Many states and jurisdictions specify parameters for how data should be collected. Regardless, most traffic stop data sets are inherently problematic, containing myriad variables recorded with missing entries. Missingness in traffic stop data can impede the ability to obtain analysis results. We use missing race information to perform a sensitivity analysis on measures of racial disparity. In particular, we ask questions about how the analysis results would be different if the missing race observations had been all White drivers or all Black drivers. We show that calculations can fundamentally shift depending on the assumptions made about the observations for which the race variable has not been recorded. That is, an initial analysis showing no racial profiling can potentially demonstrate substantial racial profiling, depending on the augmented race value for those observations that were originally missing data for the race variable.

---

## 1. Introduction

Every year, more than 20 million traffic stops are conducted in the United States, representing the most common way in which drivers interact with the police (Davis et al., 2018). Unfortunately, not all traffic stops are conducted equally—in a 2019 survey by the Pew Research Center, 59% of Black men and 31% of Black women say that they have been unfairly stopped by the police (Anderson, 2020). The relationship between belonging to a minority race, such as Black, Latine, and/or Asian Pacific Islander Desi American, and experiencing discriminatory policing is both historical and happening in the present day. Since the late 1990s, concern over racial profiling has led to federal and state mandates requiring the collection of traffic stop data (Russell, 2001).

The Stanford Open Policing Project (SOPP),[1] whose data are the focus of our study, has several dozen data sets representing over 100 million separate traffic stops (Pierson et al., 2020).[2] In tandem with the increasing availability of policing data is the growing interdisciplinary scholarship analyzing such data for evidence of racial profiling (Baumgartner et al., 2017; Grogger & Ridgeway, 2006; Knox et al., 2020; Pierson et al., 2020; Smith & Petrocelli, 2001). The traffic stops literature has developed a range of methods designed to measure racial profiling even in the setting of outcome-dependent sampling, as in the outcome test and the veil of darkness; more recent work frames discriminatory policing as a causal inference problem.

This project reanalyzes two methods of quantifying racially discriminatory policing motivated by concerns of non-ignorable missing data in the original data sets. The motivation is simple: if traffic stop data is meant to hold police officers accountable, but police officers are recording the data, can we really trust the data? To the best of our knowledge, when there is missing data in a covariate or outcome of the model, scholars generally opt to remove those observations as a simple preprocessing step.

Chanin and Welsh's (2021) review of San Diego Police Department traffic stop data point to the lack of meaningful consideration of data quality in existing traffic stop literature: of the 100 papers they reviewed, the authors find only 19 papers that address missingness and/or recording rates (Chanin & Welsh, 2021, p. 6). As one may expect, "To an officer who does not see driver race as affecting their own decision-making ... data collection only redounds to their detriment" (Chanin & Welsh, 2021, p. 9).

Motivated by the qualitative evidence of non-ignorable missingness, we conduct a missing data analysis that is organized as follows. Section 2 further contextualizes the missing data problem in the traffic stop setting and defines the different kinds of missingness. Section 3 introduces two metrics for the conditional missingness and numerous visualizations that further diagnose the problem of non-ignorable missingness. Then, we perform two sensitivity analyses in Section 4. The first examines the impact of missing data on the outcome test, a simple descriptive benchmark of the race effect on search from Pierson et al. (2020). The second sensitivity analysis is applied to the causal mediation framework developed by Knox et al. (2020) that bounds the average treatment effect (ATE) of race on the use of force during a pedestrian police stop. Our method, summarized in Figure 1, can be applied to data sets and statistics in different domains beyond policing.

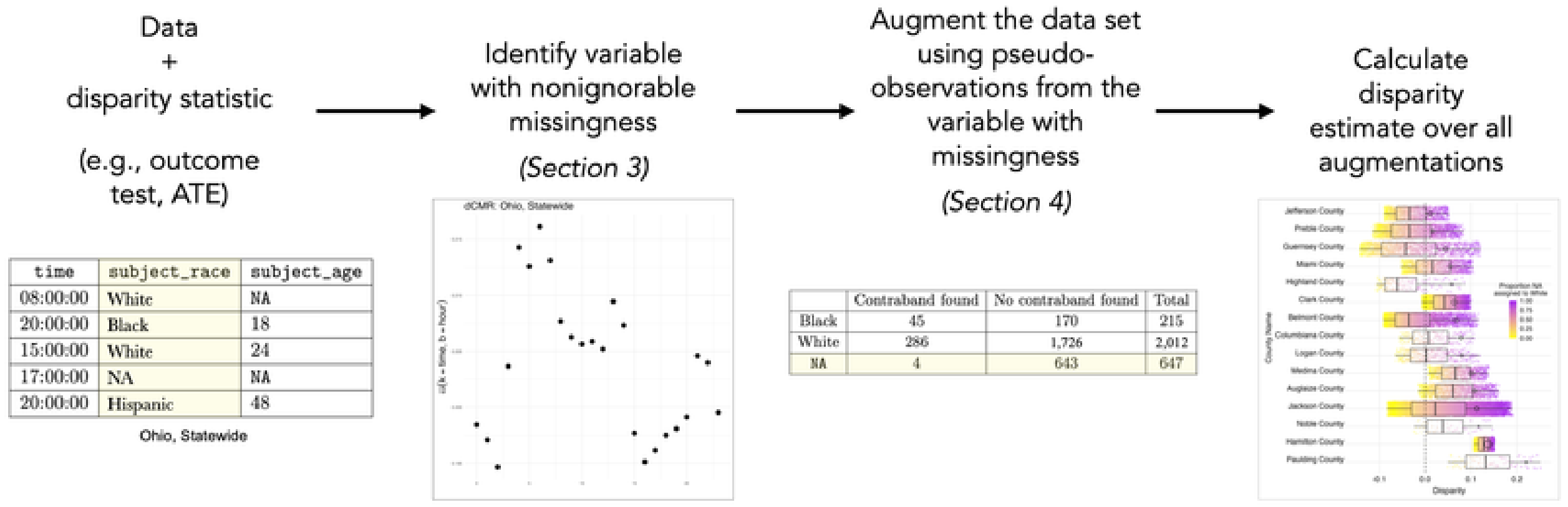

**Figure 1. A flow chart for our complete sensitivity analysis on the Ohio, statewide data from the Stanford Open Policing Project**. First, choose the disparity statistic; second, identify which variable(s) have non-ignorable missingness; third, augment the data set using pseudo-observations; fourth, calculate disparity estimate over all augmentations. See Section 4.1.1 and Section 4.2.1 for details on the process for the outcome test and average treatment effect (ATE) bounds, respectively. `NA` values are missing observations.

## 2. Missingness

The problem of missing data is present across multiple disciplines, like health care (Hunt et al., 2021; Kaambwa et al., 2012; Ward et al., 2020; Zhang et al., 2023), natural disasters (Jones et al., 2022), psychology (Gomer & Yuan, 2023), agriculture (Robbins et al., 2013), epidemiology (Allotey & Harel, 2019; Balzer et al., 2020; Tsiampalis & Panagiotakos, 2020), and crime (Blasco, 2021; Edwards et al., 2017; Stockton et al., 2024). Some of them derived methods of imputation (Blasco, 2021; Bounthavong et al., 2015; Robbins et al., 2013; Stockton et al., 2024), while others provided innovative approaches to combining information across different missingness mechanisms (Gomer & Yuan, 2023; Ward et al., 2020). A few of the papers highlight sensitivity analyses (Allotey & Harel, 2019; Stockton et al., 2024) to obtain a broader view of the missingness landscape, but their work does not use the missing data in the same way as ours does; in particular, they assume the form of the missingness mechanism (e.g., missing at random [MAR]) and we make no such a priori assumptions.

One particular study by Lo et al. (2019) on clinical trial data is closely related to our study. They describe the missingness mechanism in the pharmaceutical setting as “largely related to the post-study reporting of clinical trial data” indicating that, as in traffic stop data, there are incentives to obscure the information. By imputing missing data instead of discarding data with missing values, Lo et al. (2019) achieved superior performance in predicting successful drug approval.

In a pointed summary, Meng (2024) affirms that the

> often fatal problem is that by only learning from the cases where data are complete, one is only learning the patterns of these cases, which are systematically different from those with incomplete data when the reason for incompleteness is related to the outcome we care about.

We expand on the ideas of Meng (2024) within the context of the SOPP. Part of the motivation for our deep dive into missing data associated with traffic stops is due to policies and laws that require officers to guess at the age, race, and other demographic characteristics of the driver who has been stopped. Guessing at someone's age or race (and sometimes choosing not to guess) creates complete cases that are systematically different from the incomplete cases (e.g., see Figure 2).

Figure 2 shows two examples of some of the unusual structures in the data: in North Carolina, the age of stopped drivers has peaks in increments of five, which probably reflects some policy of officers recording perceived age instead of actual age. In Seattle, the race for stopped drivers is almost always unrecorded. Note that the lack of race data makes the Seattle SOPP data set inappropriate for an analysis on racial profiling in traffic stops. The issues in North Carolina and Seattle motivate us to think carefully about using the SOPP for large-scale analyses on racial profiling in traffic stops. The problems in the data set may result from data corruption along the way, or alternatively, they may now be resolved in more current downloads of the data. However, like many others, we use the SOPP, and the two examples foreshadow the challenge of accurately comparing racial bias across data sets when the data collection schemes are so different. In Section 4, our sensitivity analyses are limited to only data sets where missing race values correspond to nonmissing information in the search and contraband found variables.

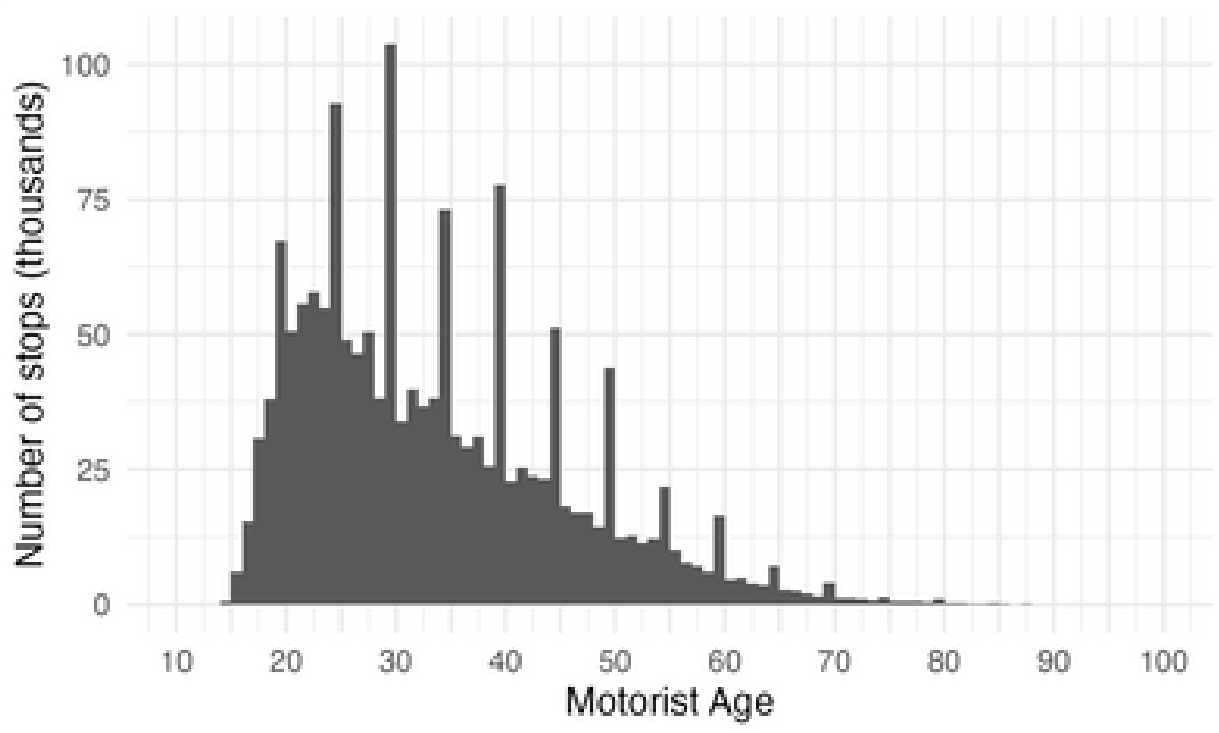

| | Black | White | NA | percent NA |
|---|---|---|---|---|
| citation | 18 | 51 | 116,661 | 99.94 |
| warning | 2 | 8 | 52,549 | 99.98 |
| arrest | 29 | 37 | 10,375 | 99.37 |
| NA | 44 | 74 | 131,508 | 99.91 |

**Figure 2. Examples of idiosyncratic data collection from the Stanford Open Policing Project database in Charlotte, North Carolina (left) and Seattle, Washington (right).** In Charlotte, the age distribution of stopped motorists has peaks at increments of five. In Seattle, the race of motorists is almost never recorded. `NA` values are missing observations.

Consider the following quotes from three different assembly bills (boldface ours).

- CA Assembly Bill No. 953 ("Racial and Identity Profiling Act of 2015", 2016)

> The perceived race or ethnicity, gender, and approximate age of the person stopped, provided that the identification of these characteristics shall be **based on the observation and perception of the peace officer** making the stop, and the information shall not be requested from the person stopped. For motor vehicle stops, this paragraph only applies to the driver, unless any actions specified under paragraph (7) apply in relation to a passenger, in which case the characteristics specified in this paragraph shall also be reported for him or her.

- New York Assembly Bill 3949 (“Prohibition on Racial and Ethnic Profiling Act”, 2015)

> [E]ach law enforcement agency shall, using a form to be determined by the division, record and retain the following information with respect to law enforcement officers employed by such agency ... the characteristics of race, color, ethnicity, gender and age of each such person, provided the identification of such characteristics shall be **based on the observation and perception of the officer** responsible for reporting the stop and the information shall not be required to be provided by the person stopped[.]

- Tennessee House Bill 2167 (“An Act to amend Tennessee Code Annotated, Title 4, Chapter 21; Title 38, Chapter 1, Part 5 and Title 49, relative to racial profiling,” 2020)

> Characteristics of race, color, ethnicity, gender, religion, and age of any person stopped for a traffic violation. The identification of such characteristics must be **based on the observation and perception of the law enforcement officer** responsible for reporting the stop, and the person stopped will not be required to provide the information.

One notable trend of these policies is that usually the race, gender identity, and age is recorded based on officers’ perception. The SOPP’s own analysis of its data points out this detail (Pierson et al., 2020). We can see evidence of ‘perceived age’ in how the age density plots of North Carolina data sets spike at multiples of five (see Figure 2), but we can also see evidence of officer perceptions in the extensive exclusion of both nonbinary gender identities and mixed-race identities and the common inclusion of Hispanic, which is an ethnicity, as a level for race. Another important feature of such state mandates are the different variables required for each state. To assess larger trends of missingness, we are able to compare data quality and missing data across data sets with myriad different variables, but data sets that collect neither search nor contraband found (or other acts of force) are outside the scope of most statistical models testing for discriminatory policing.

Our work differentiates two distinct types of missingness: (1) missing data and (2) missing potential outcomes. Missing data occur when observed data are incomplete, with values unrecorded. Missing data can be categorized as missing completely at random (MCAR), MAR, or missing not at random (MNAR); complete definitions of the terms are given in Section 2.1. In contrast, missing potential outcomes arise inherently within the framework of causal inference, where for each unit, only the outcome corresponding to the treatment received is observed, while counterfactual outcomes remain unobserved.

In a comprehensive assessment of the current missingness literature, Mitra et al. (2023) point out how "literature on causal inference with missing data is very scarce, and it is particularly challenging when both input variables and potential outcomes are systematically missing." The primary focus of this project is about non-MCAR missingness—diagnosing it and observing how it can impact downstream analyses, including causal inference methods. In order to separate these two forms of missingness clearly, we define the following terms.

## 2.1. Definitions of Missingness

### 2.1.1. Missing Data

Missing data refers to an incomplete observed data set, that is, data that was intended to have been collected but was not. Missing data is common in all types of data. In the context of policing, because much of the data are self-reported, it is not surprising that there would be extensive missing data. We refer the reader to Meng (2000), who gives an excellent tour through details of missingness, nonresponse bias, and modeling of real-life missing data mechanisms. In order to describe the missing data mechanisms, we rely on the characterizations from Little and Rubin (2020):

- **MCAR:** If missing data do not depend on the values of the data, missing or observed, that is, the probability of missing is the same for all levels of each variable, the data are called missing completely at random (Little & Rubin, 2020). Note that the definition implies that the missing data cannot be related to the research question, here, the race effect.
- **MAR:** If the missing data depend on a particular variable only through the observed components of that variable, the missingness mechanism is then called missing at random (Little & Rubin, 2020). Usually, the missingness associated with MAR data has a probability of being missing, which is a function of other variables in the data set that are completely observed.
- **MNAR:** We refer to data as missing not at random if the distribution of the missingness depends on the value of the missing components themselves (Little & Rubin, 2020) or an external unrecorded variable. The probability of data being missing is related to what the missing values would have been had we observed them. More recent work has extended MNAR to include any missing data that is not MCAR or MAR, that is, when the function describing the probability of missing data is unknown (Lo et al., 2019).

We note that MCAR and MAR are ignorable missingness. Estimates computed on MCAR/MAR data will typically be more variable (due to smaller sample sizes), but no more biased than estimates computed on complete data. Additionally, ignorable missingness allows for the possibility of performing analyses using only the observed data without introducing bias, but they require appropriate methods for analysis. The missingness due to MNAR is nonignorable, as estimates will be unreliable and potentially quite biased. We emphasize that 'without introducing bias' is the key to ignorable missingness (without the phrase, the idea is meaningless because one could always perform an analysis). Importantly, in order for the missing data mechanism to be truly ignorable, the parameter(s) in the model describing the data should have no tie(s) to the parameter(s) that regulate(s) the missing data mechanism (Rubin, 1976).

### 2.1.2. Missing Potential Outcomes

While much work has been done to estimate the race effect in traffic stops (see Section 1), the task of identifying race as a *causal* mechanism is much harder than estimating the race effect. While certainly race is an individual characteristic, we follow Knox et al. (2020) who consider race to be an aspect of the police-driver encounter rather than an aspect of the driver's identity. They say, "The manipulation of race is conceptualized as the counterfactual substitution of an individual with a different racial identity into the encounter, while holding the encounter's objective context—location, time of day, criminal activity, etc.—fixed. In other words, the 'treatment' in this case is the entire 'bundle of sticks' encapsulating the race of the civilian—including, for example, skin tone, dialect, and clothing" (Knox et al., 2020, p. 621).

In Figure 3, we extend the directed acyclic graph (DAG) from Knox et al. (2020) to illustrate the causal ordering of variables (and missing potential outcomes) in the SOPP data sets along with the missing data aspects of the variable relationships (indicated by $R$ in the graph); the Knox et al. (2020) DAG is provided by the red lines in Figure 3 (note that they include race in $X_1$ because their assumption is that all variables are complete). In their work, Knox et al. (2020) assume that any missing data (across all variables) is MCAR. One of their key assumptions is "Mandatory Reporting" explained as "encounters that escalate to the use of force also trigger a reporting requirement and are, therefore, observed in administrative data" (Knox et al., 2020, p. 625). Indeed, the work of Knox et al. (2020) is about the counterfactual of not having been stopped, a missing potential outcome (and not about missing data). We challenge their assumption (of MCAR) and build on their work (of missing potential outcomes) to incorporate data that is missing not completely at random.

Based on the ideas of Moreno-Betancur et al. (2018) (missing data) and Knox et al. (2020) (missing potential outcomes), we create a DAG model that incorporates both missing data and missing potential outcomes in the context of traffic stops (see Figue 3). The variables and their causal ordering are justified in more detail in Appendix A (in particular, Table A1 describes each connecting arrow), but importantly, in our DAG, race is a variable in the set $X_2$ (covariates with missing data), which can causally determine not only the traffic stop (mediator $M$) and the outcome ($Y$) (e.g., being searched) but also the missing data aspect of all three variables (e.g., race ($R_{X_2}$), stop ($R_M$), search ($R_Y$)).

The arrows from $X_2$ (missing data covariates) to the missingness indicators $R_{X_2}$ suggest that the missingness mechanism is MNAR (as described in Section 2.1). Additionally, the DAG in Figure 3 allows for missing data to be affected by the value of the outcome $Y$ (e.g., whether an individual was searched affects missingness in records of their stop). The DAG in Figure 3 is similar to DAG J in Moreno-Betancur et al. (2018, Figure 2), who conclude that it is not possible to recover the relationship between the outcome and the treatment in such an MNAR scenario. They say (our work is analogous to their m-DAG J)(Moreno-Betancur et al., 2018, p. 2711):

> Both m-DAG E and m-DAG J appear plausible. In both cases, the proportion exposed is nonrecoverable and sensitivity analyses would be required ... With m-DAG E, the regression-adjusted exposure-outcome association can be unbiasedly estimated using common methods, but sensitivity analyses would be required with m-DAG J.

Because the relationship is unrecoverable, we use a sensitivity analysis to provide an estimate of the range of possible values of our estimates of interest (see Section 4.2).

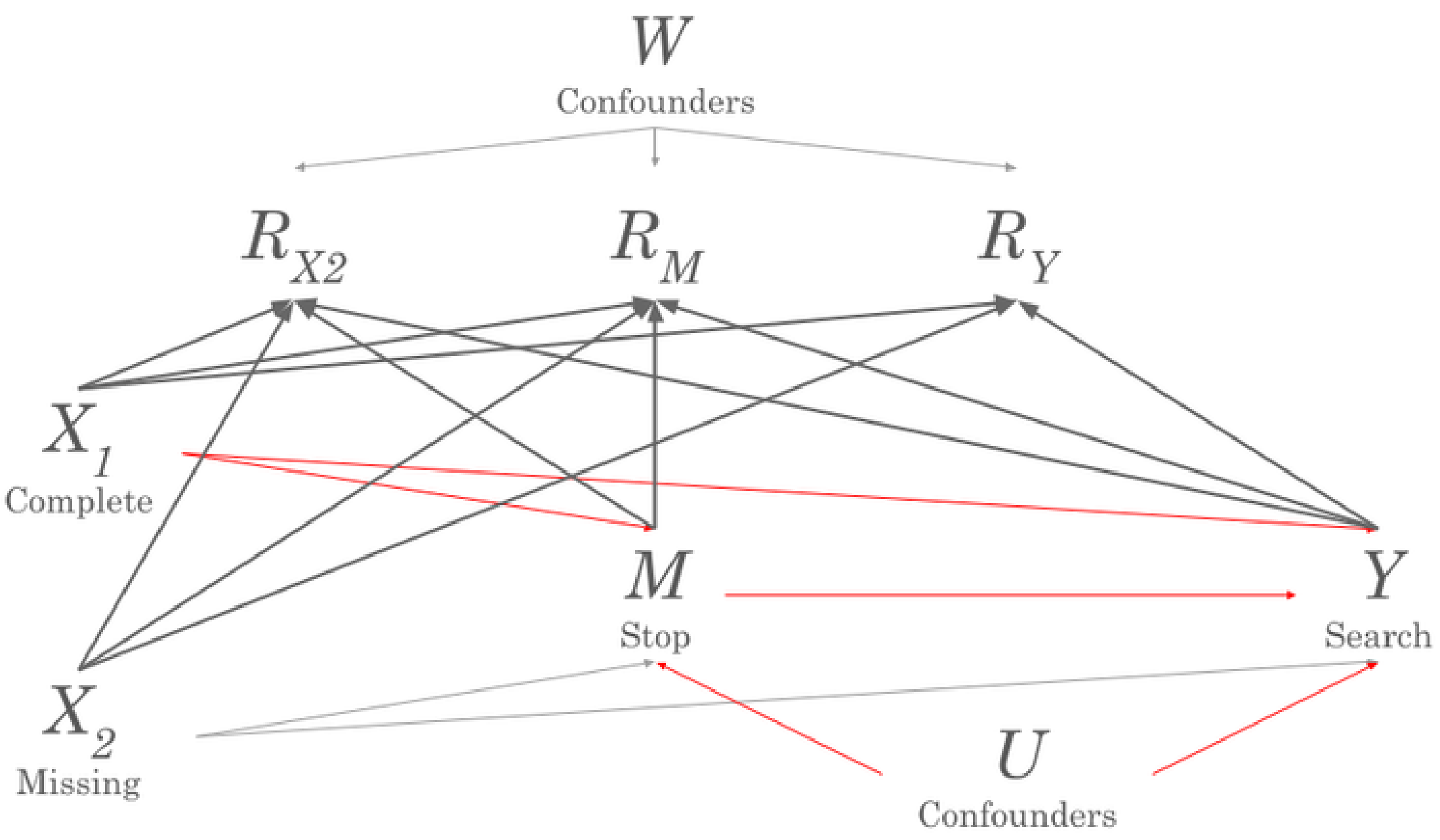

**Figure 3.** Extending the directed acrylic graph (DAG) in Knox et al. (2020) to incorporate missing data. Bold arrows depict causal mechanisms that lead to missingness. The presented graph is similar to DAG J in Moreno-Betancur et al. (2018, Figure 2) suggesting nonrecoverability and necessitating sensitivity analysis. $X_1$ represents completely observed covariates like time and day. $X_2$ represents covariates with missing information, like age and race. The Knox et al. (2020) DAG is provided by the red lines.

# 3. Diagnosing Non-MCAR

## 3.1. About the Data

The SOPP consists of 88 data sets (see Table B2) of pedestrian and vehicle stops. Each police department and highway patrol has its own idiosyncratic data collection—the years of data collection, number of observations, and number of variables are all different. To facilitate comparisons across data sets that record different numbers of variables, we narrow down our analysis to only include 20 of the most commonly recorded variables among the 88 data sets. The variables are listed in Table B1 in Appendix B. Using a subset of variables ameliorates unfair comparisons across data sets—for example, a data set with large amounts of missing values for vehicle color is not as problematic as a data set with large amounts of missing values in subject race. For each data set, we only analyze variables from the list of 20. Domain knowledge provides another way to narrow down the list of relevant variables.

## 3.2. Measuring Missingness

We begin with a toy example of a conditional missingness distribution. Consider the following five rows from vehicular stops in Chicago, Illinois, on March 14, 2016, in Table 1. Assume only five traffic stops occurred that day; also assume there are four variables recorded: date, time, race, and age.

**Table 1. Toy example of five observations from the Chicago, Illinois, data set. `NA` indicates a missing observation.**

| `date` | `time` | `subject_race` | `subject_age` |
|---|---|---|---|
| 2016-03-14 | 08:00:00 | White | `NA` |
| 2016-03-14 | 20:00:00 | Black | 18 |
| 2016-03-14 | 15:00:00 | White | 24 |
| 2016-03-14 | 17:00:00 | `NA` | `NA` |
| 2016-03-14 | 20:00:00 | Hispanic | 48 |

By glancing at each column of the table, we can quickly compute the amount of missingness in each variable when conditioning on the date being March 14, 2016: the variable time has no missingness; subject race has a 20% missingness rate; and subject age has a 40% missingness race on that particular day. Then, we could repeat these calculations for each unique value of date in the data set, and we would recover the empirical date-conditional missingness distributions, one for each of the three variables time, race, and age.

Moreover, we could aggregate the three conditional missingness distributions to produce a data set-level conditional missingness distribution on date. In Table 1, the data set-level missingness on March 14, 2016, is 20% (the average of each variable-specific missingness percentage.)

The introductory example can be formalized as follows. We first identified a conditioning variable (date). Next, we partitioned the conditioning variable into equal-sized bins by day, although we could have chosen something like week or month. Finally, we computed two types of empirical conditional missingness distributions. The first type was specific to the variable: we looked at the percent of `NA` values (where `NA` indicates missing observations) in each variable and bin. The second type provides a data set-level measure of the missingness by averaging the variable-specific empirical conditional missingness distribution. Now, we formalize the strategy into notation.

Let $X$ denote one of the 88 traffic stop data sets with $n$ observations and $m \leq 20$ variables.[3] We will denote $X_{ij}$ as the $(i, j)$ entry of the data set; $X_{\cdot j}$ as the $j$-th column; and $X_{i\cdot}$ as the $i$-th row. The binary missingness matrix $\omega$ is defined entry-wise as $\omega_{ij} = \mathbb{1}\{X_{ij} = \texttt{NA}\}$.

The distributions of $\omega_{ij}$ are unknown, so it is impossible to ask if $\omega \perp X$, or MCAR, holds. Instead, we study the conditional missingness $\omega_{ij} \mid X_{ik}$, where $X_{\cdot j}$ may be some sensitive attribute and $X_{\cdot k}$ is the *conditioning variable*, a neutral and continuous variable with naturally low levels of missingness (e.g., date). Our strategy cannot be directly used to test for MNAR. Under MNAR, the conditional missingness is dependent on something unobserved, such as the missing data itself. For the traffic stop data and other real-world data sets, natural choices of the conditioning variable $k$ include time, date, latitude, and longitude, all of which usually have low levels of missingness and are automatically recorded.

After selecting a conditioning variable, we partition the conditioning variable into $B$ ordered bins, which we denote as $b \in \{1, 2, \ldots, B\}$. We write the corresponding index set of a bin $b$ as $I_b$. There is usually a natural choice of bin—if using date as a conditioning variable, $b$ can be on the units of weeks or days; if using time as a conditioning variable, $b$ can be on the unit of hour. Next, for each $b$, we can compute the percent of missingness for observations in that bin $I_b$. The percent of missingness can be calculated either for a single sensitive attribute $X_{\cdot j}$ (e.g., the percent of missingness in race on a particular day) or across all $m - 1$ variables excluding the conditioning variable (e.g., the percent of missingness on a particular day, aggregated across all variables).

We define these two measurements of conditional missingness. The first is for a single covariate; the second aggregates over $m - 1$ covariates.[4]

- The conditional missingness rate (CMR) of covariate $j$ given some conditioning variable $k$ taking values in bin $b$ is the average amount of missingness in covariate $k$ for observations in bin $b$.
$$\overline{\omega}_j(k, b) = \frac{1}{|I_b|} \sum_{i \in I_b} \omega_{ij} \quad \text{where} \quad I_b = \{i \in [n] \mid X_{ik} \in b\}. \quad (3.1)$$
- The data set conditional missingness rate (dCMR) given some conditioning variable $k$ taking values in bin $b$ is the average amount of missingness in the entire data set for observations in bin $b$.
$$\overline{\omega}(k, b) = \frac{1}{m - 1} \sum_{j \neq k} \overline{\omega}_j(k, b). \quad (3.2)$$

The purpose of dCMR is to see the relationship between missing values and some conditioning variable. While we do not formally test the independence of the conditioning variable and the missingness (as one might do if they were formally testing for non-MCAR), we will later compute the maximal correlation between the conditioning variable and the dCMRs. In the plots below, the calculation of dCMR is done by averaging over the set of variables (when available) given in Table B1 for each data set.

The maximal correlation measures nonparametric dependence between two variables (Breiman & Friedman, 1985). Maximal correlation does not assume a linear or monotonic relationship between two variables (unlike the Pearson correlation or Spearman's rank correlation, respectively). The maximal correlation finds transformations of each variable that maximize the population-level correlation. While it can be used to formally test for independence, we use maximal correlation as a summary statistic to measure the dependence of two variables. A high value of maximal correlation indicates that there is a (nonlinear) relationship between the dCMR and the variable of interest (see, e.g., Figures 4–8 where all of the relationships have quite high maximal correlation). We are not able to use standard tests of MAR [e.g., Little's test, Little, 1988] because our data are not normally distributed. The maximal correlation gives us a sense of dependency between missingness and other variables without using a formal test. We use the R package `acepack`, which implements the alternating conditional expectations algorithm to compute the maximal correlation (Spector et al., 2025).

When auditing a large number of data sets and variables for non-MCAR, the maximal correlation offers a quick way to understand the missingness patterns at a high level. If most data sets have low maximal correlation, then an analyst might not need to manually check each data set for missingness patterns.

## 3.3. Notable dCMR Trends Across the SOPP

Here we provide a high-level audit of the entire 88 data sets (see Table B2) in the SOPP when conditioning on date and binned by week. Date as a conditioning variable is especially natural because all 88 data sets record date (although some data sets have some missingness in date).

We reviewed the dCMRs for all data sets and selected the six plots in Figure 4 as representative of common trends within the SOPP data sets (all of which seem to be not MCAR). The problematic relationships between dCMR and date (by week) include piecewise constant, changes in variance, and nonmonotonic trends across time. The information within the data itself does not provide reasons for the patterns; they could be due to procedural changes, changes in how the data were collected, or changes in missingness for particular subsets of variables. If the research goal is to truly understand the exact missingness patterns, we recommend a further dive into the CMR plots for individual variables. Note, however, that our work has found no strong trends between the additional variables and the variables of interest in our work (i.e., race and outcome measures like searched and contraband found). The dCMR (missingness on the *y*-axis) for each of the plots is on a different scale. We let the scales be free in order to focus on the trends in the missingness patterns.

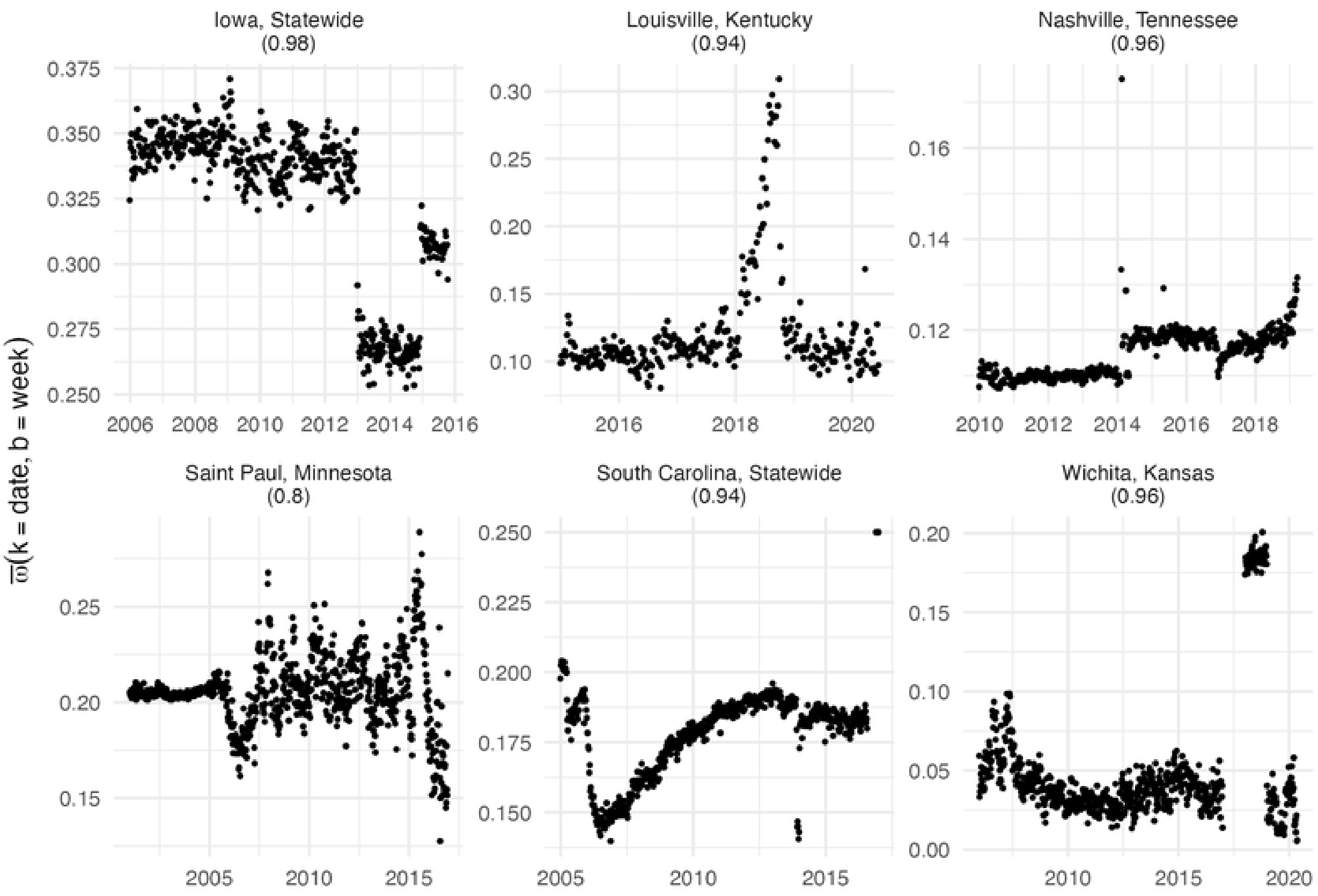

**Figure 4. The data set conditional missingness rate (dCMR) of six selected data sets conditioned on the date (binned into weeks).** The number given in parentheses is the maximal correlation between dCMR and date. The data sets demonstrate heterogeneity over time: level changes, changes in variance, and nonmonotonic changes.

In Figure 5 we demonstrate the usefulness of the maximal correlation by selecting the five plots corresponding to the minimum maximal correlation, the maximal correlation quartiles (25th, 50th, and 75th percentiles), and the maximum maximal correlation, providing a high-level summary of the extent of dependence between missingness and date. One might argue that we cherry-picked the data sets in Figure 4. However, the maximal correlation values in Figure 4 are around the 50th and 75th quantiles of the maximal correlation distribution (seen in Figure 5), indicating that the data sets described in Figure 4 are not extreme cases compared to the overall trends we see across all of the data sets. Note that the dCMR (missingness on the *y*-axis) for each of the plots is on a different scale. We let the scales be free in order to focus on the trends in the missingness patterns.

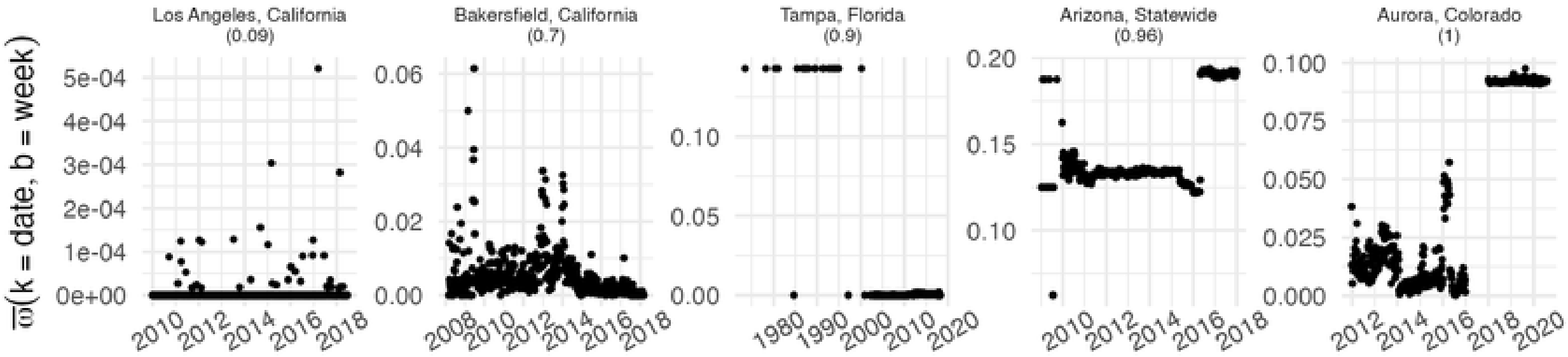

**Figure 5.** The data set conditional missingness rate (dCMR) of five data sets conditional on date (binned into weeks) corresponding to the minimum, quartiles (25th, 50th, and 75th percentiles), and maximum maximal correlation arranged from left to right; in parentheses is the maximal correlation between dCMR and date.

## 3.4. Auditing Eight Data Sets for Non-MCAR

In Section 4, we consider how non-MCAR bias can be assessed in models of discrimination that rely on three covariates: race, whether or not a search was conducted, and whether or not contraband was found. Thus, here we tailor our analysis to the eight data sets that have at least 10% missingness in at least one of race, search, or contraband (with all three of the variables recorded for the data set). The eight data sets consist of seven statewide data sets and only one city-level data set: California, Colorado, Chicago, Illinois, Maryland, New Jersey, Ohio, Washington, and Wisconsin. Of the eight data sets, eight record the variable date, six record the variable time, and four record the variables' latitude and longitude. The dCMR is computed using the 20 variables described in Table_B1.

### 3.4.1. Temporal Trends

First, we examine how missingness correlates with date in Figure 6. Although the scale of the *y*-axis is small for many of these plots, we clearly see nonrandom missingness trends over time. Three data sets (California, Colorado, and Washington) have increasing missingness over time; two (Chicago and New Jersey) have decreasing; and three (Maryland, Ohio, and Wisconsin) have trends that are roughly equal over time, with each state having a handful of years that are clearly outliers with respect to the missingness. As with Figure 4, we cannot know the reason for the missingness patterns. A deeper dive into the CMR plots for individual variables could shed light into some of the relationships, however, we have found that none of the variables are strongly associated with the missingness trends connecting race and outcome variables (like search and contraband found).

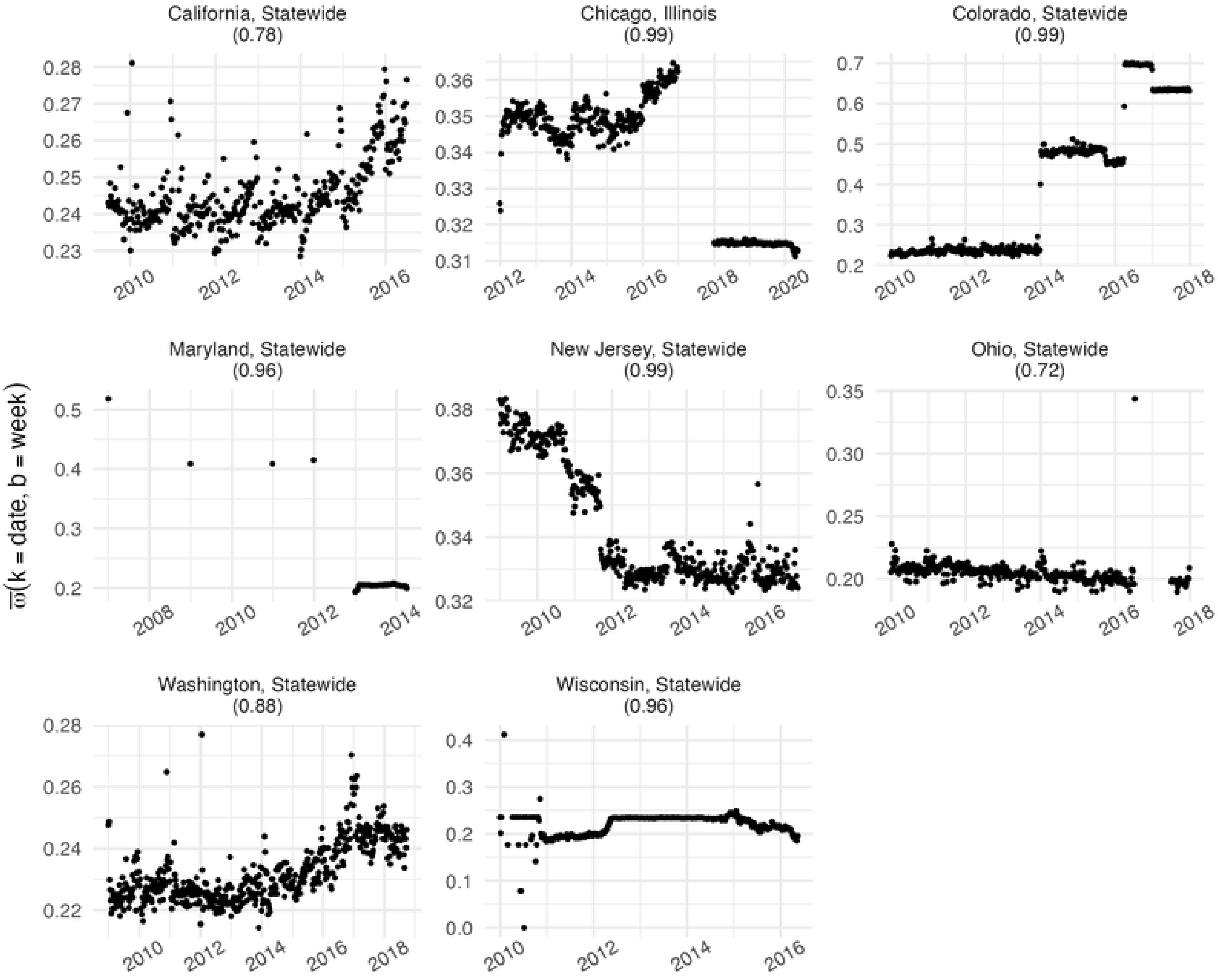

**Figure 6. The data set conditional missingness rate of eight data sets conditional on date (binned into weeks).** The maximal correlation for each data set is included in parentheses.

When considered across time (binned by hour), the dCMRs also exhibit dependence with the conditioning variable in Figure 7. In four data sets (Chicago, New Jersey, Ohio, and Wisconsin), missingness peaks in the early mornings; in two data sets (Maryland and Washington State), missingness has two local peaks—one in the early morning, and one later in the day. However, we point out that the range of dCMR missingness is quite small, and the peaks in missingness may have to do with external factors.

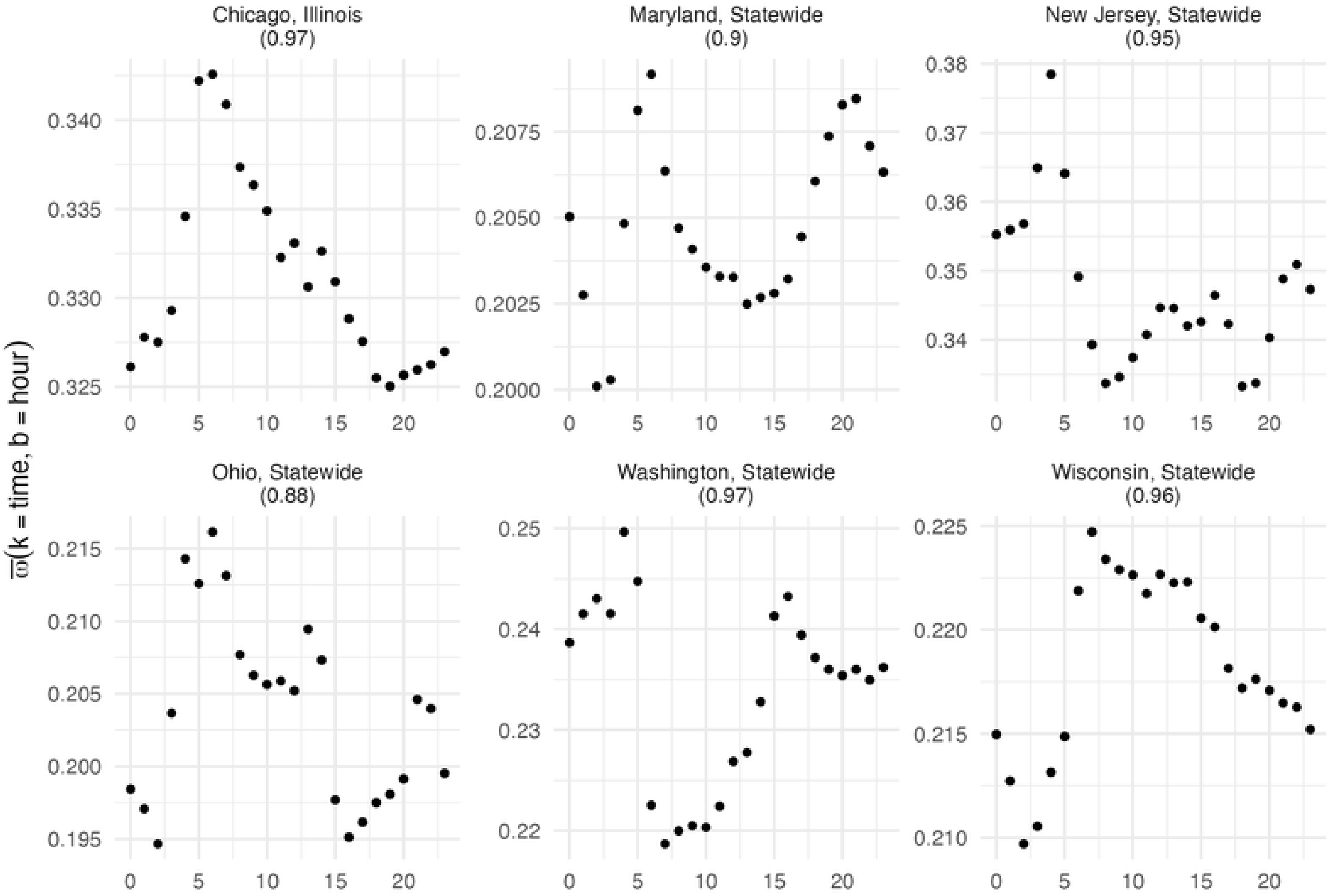

**Figure 7. The data set conditional missingness rate of six data sets given the time (binned into hours).** The maximal correlation for each data set is included in parentheses.

### 3.4.2. Spatial Trends

When we condition on latitude and longitude, we can plot each bin (6-digit geohash) with the $x$-axis as longitude and the $y$-axis as latitude. Then, the dCMR can be color-coded. For the three state-level data sets in Figure 8, the dCMR plot on latitude and longitude naturally recovers state highways (although it is not a perfect representation due to some highway patrols working primarily on highways [WA] and some patrolling side streets as well [OH]). The maximal correlation here is an average of the maximal correlation between latitude and dCMR and between longitude and dCMR. The maximal correlation between location and missingness in statewide data sets is low, with only Wisconsin state having maximal correlation above 0.1. Near city centers like Minneapolis toward the west and Milwaukee in the east, the dCMR is slightly higher than its surroundings. Meanwhile, in Chicago, one can clearly see the higher levels of missingness in the city center and lower missingness in the suburbs.

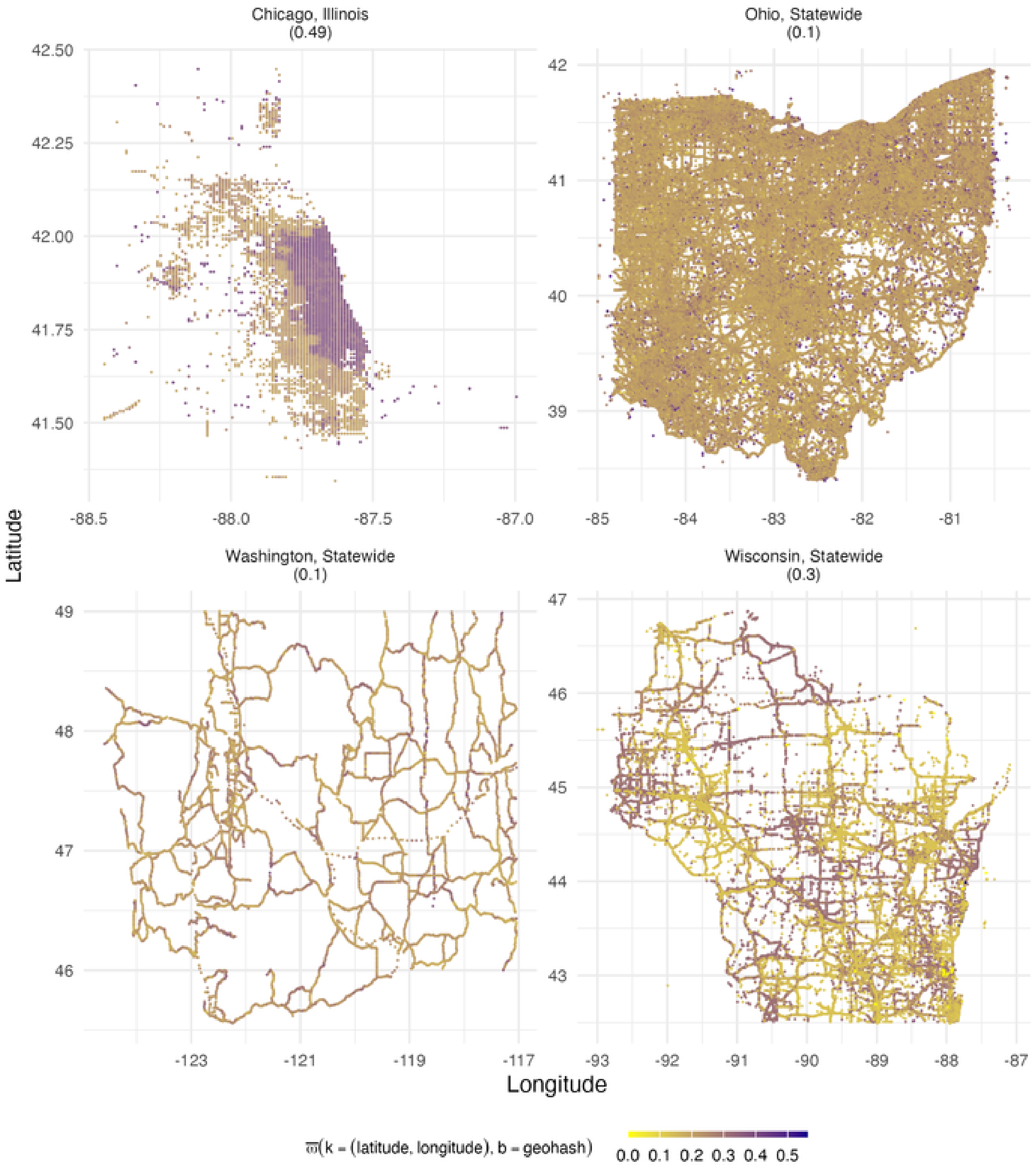

**Figure 8. The data set conditional missingness rate (dCMR) of four data sets given the latitude and longitude (binned into 6-digit geohashes)**. The maximal correlation (average of the maximal correlation of latitude & dCMR and longitude & dCMR) for each data set is included in parentheses.

## 4. Assessing Sensitivity of Missingness

Using the SOPP, we analyze several data sets to broadly explore differential missingness. As a way to investigate the impact of the missing race data, we assign observations whose race is not measured to either the Black or White groups of individuals, randomly. Subsequently, we apply two published methods to the augmented data set in order to investigate how missingness can impact racial bias estimates. Although both published studies consider race within the framework of missing potential outcomes, neither study delves into the impact of missing data in considering the reporting of a race effect. Our sensitivity analysis furthers the literature on understanding racial discrimination in traffic policing. Our work has similarities to Meyer and Gonzalez (2024) who use a sensitivity analysis on a new test, the overlapping condition test.

## 4.1. The Outcome Test

An analysis of over 20 million traffic stops applies the outcome test (originally described in Becker [1993] applied to traffic data in Pierson et al. [2020, Figure 3]) to compare the rate of searches to the rate of successful searches in order to determine if the police officers use a lower standard for searching drivers of color (race effect). The researchers find evidence of both racial profiling and preferential searches for minority race drivers. Although the discussion in Pierson et al. (2020) focuses on standardizing and improving the quality of traffic stop data collection, it does not mention how missing data plays a role in the analyses.

Applied to the traffic data, the outcome test (Pierson et al., 2020, Figure 3) compares 'hit rates' (point estimates of the probability of recovering contraband given that a driver is searched) between Black and White drivers. The idea is that if searches of minority drivers reveal contraband at lower rates than searches of White drivers, the outcome test suggests that officers are searching minority drivers on the basis of less evidence and with lower success; that is, they are potentially racially biased.

The outcome test is not designed to identify causal mechanisms for differential hit rates; instead, it is a measure of the race effect. For example, a lower hit rate could be due to racial bias of the officers or because Black drivers are carrying around less contraband than White drivers. However, the latter reason would require that (1) the police be colorblind and (2) the police are not using any other correlated indicators related to contraband (e.g., the age of the vehicle). Our work does not delve into the shortcomings of the outcome test (e.g., inability to identify causal indicators), which have been studied in detail in prior work (Engel, 2008). Instead, we offer a sensitivity analysis that assesses the impact of missing data race information. We offer our sensitivity analysis to complement analyses such as Becker (1993), Persico and Todd (2006), Pierson et al. (2020), and Gaebler and Goel (2025) who use the outcome test in their own work.

### 4.1.1. Sensitivity Analysis of the Outcome Test

Throughout our sensitivity analysis, we use disparity as a measure of racial discrimination. Equation 4.1 provides the generic form of disparity where $p$ represents the probability of some police action (e.g., being searched). Equations 4.2 and 4.3 are specific to the outcome test (here, Section 4.1.1) and bounds on the ATE (see Section 4.2.1), respectively. It is important to note, however, that $\text{disparity}_{\text{outcome}} < 0$ indicates discrimination against Black drivers because if Black drivers have lower rates of contraband found than White drivers, they are being searched unfairly. On the other hand, $\text{disparity}_{\text{ATE}} > 0$ indicates discrimination against Black drivers because the traffic stop 'as if Black' is being searched more often than the traffic stop 'as if White.'

$$\begin{aligned} \text{disparity} &= p_{\text{Black}} - p_{\text{White}} && (4.1) \\ \text{disparity}_{\text{outcome}} &= P(\text{contraband} \mid \text{stop, search, Black}) - P(\text{contraband} \mid \text{stop, search, White}) && (4.2) \\ \text{disparity}_{\text{ATE}} &= P(\text{search} \mid \text{Black, stop as if Black}) - P(\text{search} \mid \text{White, stop as if White}) && (4.3) \end{aligned}$$

The first step of the sensitivity analysis is to calculate the *ignore NA* disparity: point estimate of the probability of recovering contraband given that a driver is searched for Black drivers minus White drivers (without considering any observations with `NA` race).

Next, we ask how the disparity would change if the observations with missing race had not been missing. For example, what would the disparity be if all the observations with missing race turned out to be Black drivers? What would the disparity be if all the observations with missing race turned out to be White drivers? Or something in between? As described below (and worked out in detail in Table 3), our sensitivity analyses create new data sets (referred to as *augmented* data) by allocating the `NA` race observations to either the Black or White group. We call the allocated `NA` race observations *pseudo-observations*; we use the word 'pseudo' because they represent actual observations from the data set, but their race has been augmented through the sensitivity analysis.

Figure 9 presents a sensitivity analysis based on allocating `NA` race drivers to either the Black or White driver groups for a random subset of counties in Ohio (the image for all Ohio counties is given in Appendix C in Figure C5. The left plot of Figure 9 presents the *ignore NA* county-level disparities. Note that all of the *ignore NA* county-level disparities are greater than zero, which indicates that there is no evidence of discrimination against Black drivers. The right plot presents all possible disparities from assigning `NA` race observations as either Black or White. That is, every possible allocation (see Table 3) of the missing race observations across the Black and White groups is considered, and the disparity is calculated on the augmented data. Each dot in the right plot represents one possible allocation of the `NA` race observations. We recognize that allocating all of the `NA` race drivers to Black can happen in exactly one way; however, allocating half of the `NA` race drivers to Black can happen in many ways (assuming each individual driver is distinct). We calculated the number of combinations, not the number of permutations because using the frequency associated with the number of permutations would imply that the `NA` race drivers were equally likely to be distributed across Black and White drivers. Instead, the sensitivity analysis allows us to see the *possible* values given any potential distribution of the `NA` drivers to Black and White. In the right plot, the *ignore NA* disparity estimators are given as open black circles, which trend more positive (less discrimination against Black drivers) than the vast majority of measurements obtained from the sensitivity analysis (possible disparities across all potential allocations of `NA` drivers to Black and White). Tables 2 and 3 detail the process of all possible allocations of `NA` drivers to the Black and White driver groups for Belmont County, Ohio (visualized in the middle of Figure 9). Table 2 provides the original data for Belmont County, Ohio. Table 3 describes the process by which the data are augmented by systematically allocating `NA` race values to either the Black or White group.

We see that incorporating missingness leads to estimates with lower disparities, indicating that incorporating the missing information tells us that more discrimination against Black drivers is happening than we had previously understood (with the *ignore NA* estimator only). The color of the point in the right panel of Figure 9 indicates the proportion of missing race observations that were allocated to the White group for calculations.

**Table 2. Original data set for Belmont County, Ohio. `NA` values are missing observations.**

| | Contraband found | No contraband found | Total | Naive hit rate |
|---|---|---|---|---|
| Black | 45 | 170 | 215 | 45/215 = 0.209 |
| White | 286 | 1,726 | 2,012 | 286/2,012 = 0.142 |
| `NA` | 4 | 643 | 647 | 4/647 = 0.006 |

**Table 3. Augment the data set by allocating `NA` race observations to Black and White groups for Belmont County, Ohio. `NA` values are missing observations.[5]**

| `NA` race contraband $\rightarrow$ B | `NA` race contraband $\rightarrow$ W | `NA` race no contraband $\rightarrow$ B | `NA` race no contraband $\rightarrow$ W | disparity (Equation 4.1) |
|---|---|---|---|---|
| Sums to 4 | | Sums to 643 | | |
| 4 | 0 | 643 | 0 | $\frac{45+4}{215+4+643} - \frac{286+0}{2,012+0+}$ |
| 3 | 1 | 643 | 0 | $\frac{45+3}{215+3+643} - \frac{286+1}{2,012+1+}$ |
| ⋮ | ⋮ | ⋮ | ⋮ | |
| 0 | 4 | 643 | 0 | $\frac{45+0}{215+0+643} - \frac{286+4}{2,012+4+}$ |
| ⋮ | ⋮ | ⋮ | ⋮ | |
| 4 | 0 | 0 | 643 | $\frac{45+4}{215+4+0} - \frac{286+0}{2,012+0+64}$ |
| ⋮ | ⋮ | ⋮ | ⋮ | |
| 0 | 4 | 0 | 643 | $\frac{45+0}{215+0+0} - \frac{286+4}{2,012+4+64}$ |

To quantify the changes from the *ignore NA* estimator, note the following, which are summarized in Table 4. Of the 87 counties in Ohio, 10 indicate discrimination against Black drivers (10 open dots to the left of the vertical dashed line at zero in Figure C5) and 77 indicate discrimination against White (77 open dots to the right of the vertical dashed line at zero in Figure C5). However, when considering the sensitivity, 71 counties could switch disparity estimate (71 points could switch sides of the vertical line at $x = 0$): *66 could switch from discrimination against White to discrimination against Black drivers*, five could switch from discrimination against Black drivers to discrimination against White drivers, and 16 counties could not switch discrimination, as measured by the sensitivity analysis.

**Table 4. Outcome test disparity estimates.** $d = p_{Black} - p_{White} > 0\ (+)$ **indicating discrimination against White drivers;** $d = p_{Black} - p_{White} < 0\ (-)$ **indicating discrimination against Black drivers.** `NA` **values are missing observations.**[6]

| Data set | Total number of countries/departments | Countries/departments with missingness | Sign of *ignore NA* disparity estimate | | Imputed county-level disparity estimate | | | |
|---|---|---|---|---|---|---|---|---|
| | | | $(-)$ | $(+)$ | Counties that could switch signs | | Counties that do not switch signs | |
| | | | | | $(-)$ to $(+)$ | $(+)$ to $(-)$ | Remains $(-)$ | Remains $(+)$ |
| OH | 89 | 87 | 10/87 | **77/87** | 5 | **66** | 5 | 11 |
| CO | 66 | 32 | 25/32 | 7/32 | 0 | 2 | 25 | 5 |
| WI | 72 | 49 | 25/49 | 24/49 | 3 | 2 | 22 | 22 |
| WA | 39 | 31 | 19/31 | 12/31 | 3 | 3 | 16 | 9 |
| MD | 223 | 60 | 34/60 | 26/60 | 0 | 2 | 34 | 24 |

Appendix C contains the same sensitivity analysis for the states described in Section 3.4. Note that we previously identified eight data sets that had missingness of at least 10% in each of the variables race, search, and contraband. However, New Jersey and California plus Chicago, Illinois, data sets do not contain relevant information for the outcome test. That is, when race is `NA`, all searches conducted and contraband found are also `NA` in New Jersey, California, and Chicago. We can get *ignore NA* estimators for each state, but we cannot repeat our sensitivity analysis for those states where missingness in race, search, and contraband happens simultaneously. The analysis of missing race data is given in full for Ohio, Colorado, Wisconsin, Washington, and Maryland highway patrols in Appendix C.

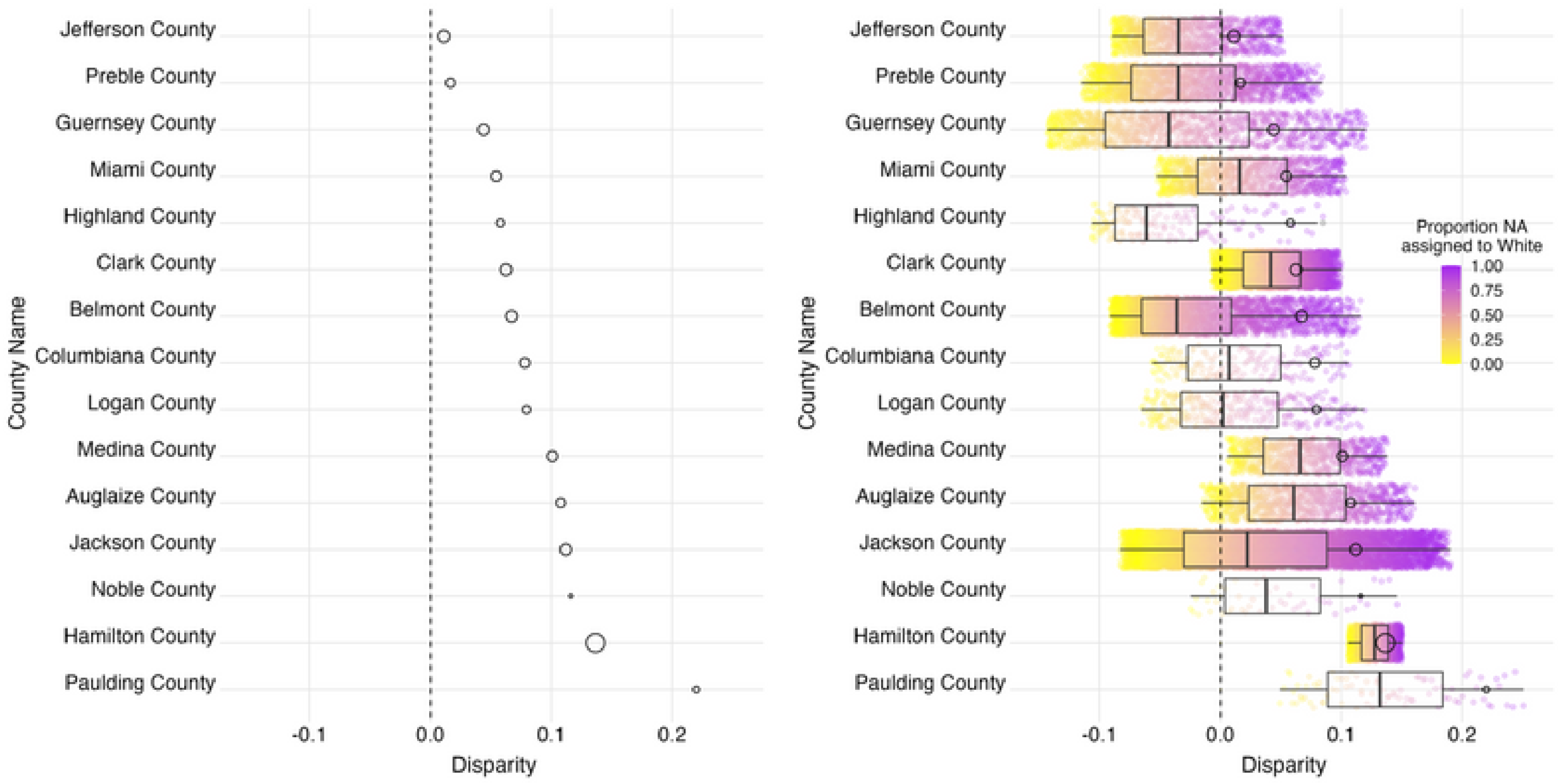

**Figure 9.** On both images, the *ignore NA* county-level disparities are given by open black circles sized by the number of total searches from that county in Ohio. The right plot presents all possible disparities from assigning `NA` race observations to either Black or White. Boxplots describe the distribution of augmented disparities across the allocations of missing race observations. In each boxplot, the black horizontal line is given as the median of the augmented disparities across all possible allocations. The color of the point in the right panel indicates the proportion of missing race observations that were allocated to the White group. `NA` values are missing observations.

## 4.2. Bounds on the Average Treatment Effect

Knox et al. (2020, p 621) conceptualize race as a causal variable (missing potential outcomes) and the "investigation of racial bias in policing as an inherently causal inquiry." Using causal inference, they derive strict bounds for the ATE (the extent to which drivers of color face greater risk of force than White drivers because of their race), given a traffic stop was made. In particular, the ATE measures the probability of being searched if a person was Black and stopped as if they were Black minus the probability of being searched if a person was White and stopped as if they were White, given that a stop was made, see Equation 4.3. They do not discuss missing data or its impact on the analyses. However, they do require the "Mandatory Reporting" assumption (that all encounters are recorded) in their derivations (an assumption that is clearly violated if some data are missing)(Knox et al., 2020, p. 625) .

Knox et al. (2020) describe the missing potential outcomes shown in Figure 3 as the triangle between $X_2$, $M$, and $Y$. We extend their work to include missing data ($R_{X_2}$, $R_X$, and $R_Y$). Because it is not possible to recover the relationship between outcome and treatment in the missing data / missing potential outcomes scenario (Moreno-Betancur et al., 2018), we provide simulated bounds to assess the impact of $R_{X_2}$, missing data.

Knox et al. (2020) derive nonparametric bounds of ATE (given a traffic stop was made) on being searched, as a function of $\rho$, the proportion of racially discriminatory stops. Note that $\rho$ is unknowable, and the Knox et al. (2020) results are provided over a range of possible $\rho$ values. As such, they are able to measure the missing potential outcomes information (in the complete data case), but they do not evaluate the impact of missing data on the missing potential outcomes analysis.

### 4.2.1. Sensitivity Analysis of the Bounds on Average Treatment Effect

To address possible non-MCAR missingness in race on ATE, we perform another sensitivity analysis where we calculate the ATE bounds when `NA` values have been ignored [equivalent to Knox et al., 2020, which are based on bounds by Horowitz and Manski, 2000] and compare with the bounds when NA values have been allocated to the Black and White groups. Consider Colorado and Washington highway patrols, as shown in Figure 10. Following Knox et al. (2020), by ignoring the NA race values, we calculate the naïve disparity and also bounds on the disparity for each of three different values of $\rho$ (the proportion of racially discriminatory stops): 0.25, 0.5, 0.75. For each of three different values of $\rho$ we calculate the bounds on the ATE, while ignoring the NA values (just as Knox et al., 2020). Those bounds are represented by the dashed vertical lines and the corresponding ribbon plots. The solid vertical lines are the Knox et al. ATE bounds at each of a different value for the proportion NA assigned to White. The red dots represent the estimated ATE (naïve estimate) at each value for the proportion NA assigned to White. As expected, a higher proportion of racially discriminatory stops naturally leads to a higher range of values of disparity.

The sensitivity of the ATE disparity estimator did not enumerate all possible combinations associated with allocating the NA race observations. Instead, to create pseudo-observations, we allocated a fixed proportion of NA race observations to White (0, 0.25, 0.50, 0.75, and 1.0), and, for each proportion, we randomly selected that proportion of the NA race drivers (and their search outcome) to assign to the White drivers and the remaining NA race drivers (and their search outcome) were assigned to the Black drivers, resulting in an augmented data set. Additionally, we did two extreme sensitivity analyses: first, we assigned all of the NA race and searched drivers to the Black driver group (and all of the NA race and nonsearched drivers to the White driver group); second, we assigned all of the NA race and searched drivers to the White driver group (and all of the NA race and nonsearched drivers to the Black driver group).

Recall that $\text{disparity}_{\text{ATE}} > 0$ indicates discrimination against Black drivers. Figure 10 (and additional figures in Appendix D) show that in most original estimates (i.e., not including NA drivers) and sensitivity analyses, the disparity indicates discrimination against Black drivers.

In many of the data sets, there are very low rates of `search = TRUE` within the NA race drivers. Additionally, the group of Black drivers is usually much smaller than the group of White drivers. Therefore, when assigning most of the NA race drivers to the Black driver group (i.e., proportion NA assigned to White is low), the Black driver group seems *less frequently searched* than *ignore NA* estimates (i.e., the discrimination against Black drivers seems lower when most of the `NA` race drivers are assigned to White).

Note that the 'extreme' cases (Figure 10 and Appendix D) seem to violate the strict bounds of Knox et al. (2020). However, the bounds from Knox et al. (2020) are only strict if there is no missingness (or if the missingness is MCAR) and are derived given the DAG of Knox et al. (2020), which is the red subset of the DAG in Figure 3. In Figure 3 and Section 3 we justified arrows from the variables to the missingness indicators that motivate our sensitivity analysis. And the sensitivity analysis demonstrates that the strict bounds of Knox et al. (2020) are not strict if the `NA` race observations are considered.

Appendix D contains the same sensitivity analysis as Figure 10 (proportion of `NA` race drivers assigned to White as well as extreme value plots) for the states described in Section 3.4. Note that we previously identified eight states that had missingness of at least 10% in one of the variables race, search, and contraband. As before, New Jersey and California highway patrol and Chicago, Illinois, data sets do not contain relevant information for the ATE sensitivity analysis. That is, when race is `NA`, all search conducted is `NA` in New Jersey, California, and Chicago. We can get estimates that ignore `NA`, but we cannot perform the sensitivity analysis.

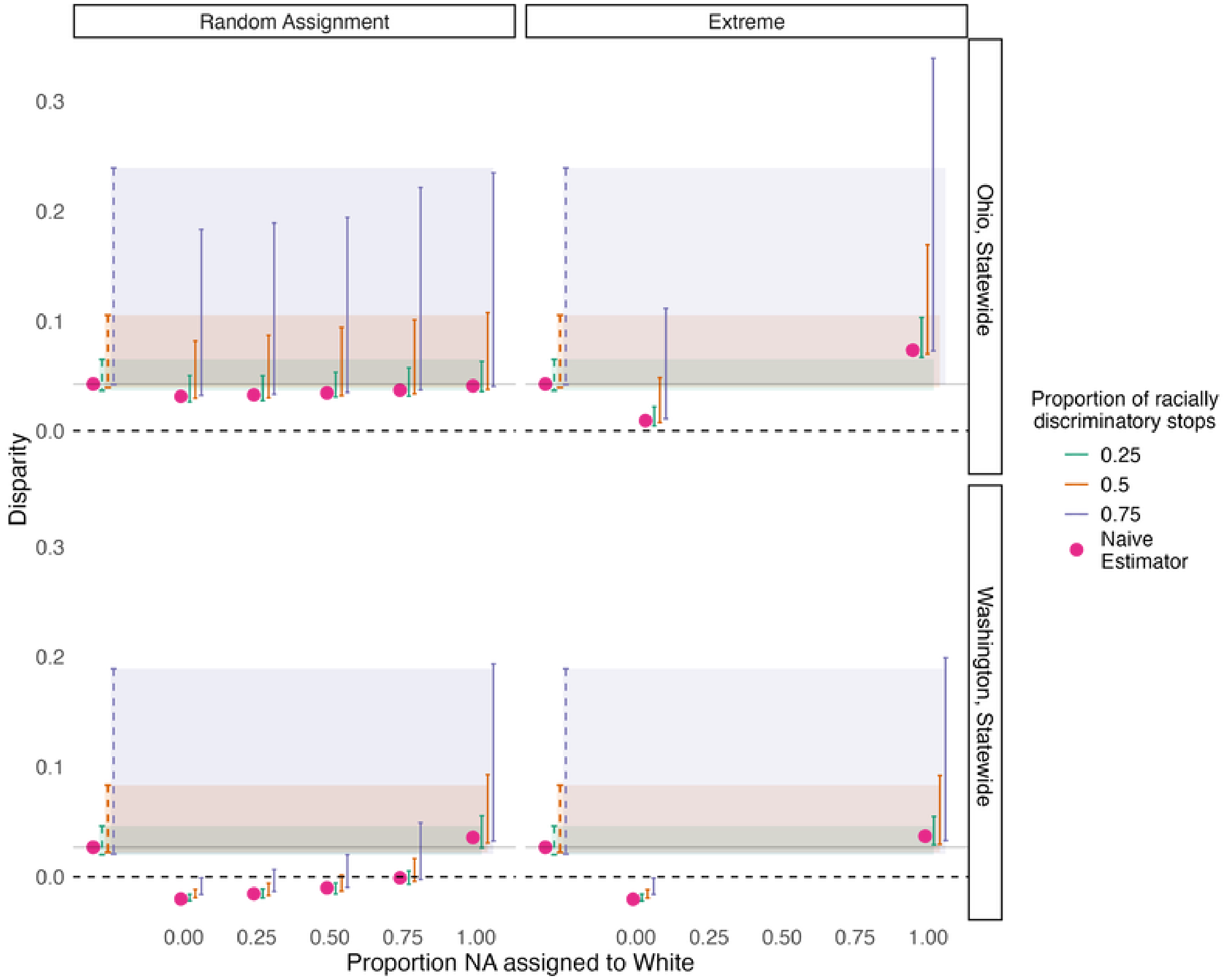

**Figure 10.** Dashed line at zero indicates no differential racial impact. Disparity is the difference in proportion of search for Black minus proportion of search for White. Disparity is measured as bounds on the search ATE, given recorded outcome, for different levels of the $\rho$ parameter (the overall proportion of racially discriminatory stops). The $x$-axis of each plot presents a different allocation of assigning `NA` race observations to either Black or White. `NA` values are missing observations.

## 4.3. Discussion of Sensitivity Analysis

The goal of our work was to investigate the impact of including **NA** race drivers in methods used to identify racial discrimination of traffic stops. That is, is the race of the driver being covered up to avoid any perception of discrimination? We were able to show that across the SOPP the data sets have quite a bit of missing information that is not MCAR. By focusing on observations that have missing race data, we were able to perform sensitivity analyses to understand the impact of missing race on two different discrimination metrics.

The outcome test looks at discrimination in rates of contraband found. Table 4 demonstrates that the **NA** race drivers, when added to the non **NA** race drivers, can change our perspective on discrimination. In particular,

Ohio shows that in most counties, it is possible for the `NA` race drivers to change the perceptions from discrimination against White drivers to discrimination against Black drivers. Other counties show some of the same effect but to a lesser degree.

The sensitivity analysis done on the metric that provides strict bounds on the ATE gave insight into discrimination based on the search variable. The *ignore NA* statistics indicated that there is some discrimination against Black drivers with respect to the search variable. In Ohio, the discrimination still exists, even in the most extreme case of the sensitivity analysis. While some other states do show that there could be a lessening of discrimination against Black drivers (when `NA` race drivers are included), the general sense is that, due to the sensitivity analysis, there is anti-Black discrimination as measured by the ATE disparity.

## 5. Discussion

As a way to objectively assess the amount of missingness in a data set, we define two metrics in Section 3.2 to quantify the distribution of missingness given some conditioning variable. Our metrics are able to identify both temporal and spatial relationships, and we point out that missingness patterns across either time or space could indicate problematic data collection habits of the peace officer.

Our work demonstrates how missing data information can be used to understand missingness trends and the sensitivity of racial disparity outcomes to missing data. We clarify the distinction between missing data and missing potential outcomes and reason that understanding both is important. Our sensitivity analysis is applied to two metrics of racial discrimination: the outcome test (Pierson et al., 2020) and bounds on the ATE (Knox et al., 2020). There are many other metrics that have been used to determine racial bias in traffic stops and could be assessed using a similar sensitivity analysis, for example, the Knowles, Persico, and Todd test (Knowles et al., 2001), the threshold test (Pierson et al., 2018, 2020; Simoiu et al., 2017), internal benchmarking (Ridgeway & MacDonald, 2010), or the hit rates test (Persico & Todd, 2006).

Note that there are methods, such as the "veil of darkness" (Grogger & Ridgeway, 2006), which have been able to directly address the problem of data reporting rates that vary by race through modeling. In particular, their model uses ratios of probabilities (of data reporting) that can vary by race but do not vary given race and time. Not every model, however, can easily integrate probabilities of reporting, and we believe that our method of running a sensitivity analysis can benefit many models that measure racial disparities in policing.

Additionally, the methods provided here are applied exclusively to traffic stop data. However, the ideas could be expanded to data of many different types. As mentioned in Section 2, we see the importance of addressing missingness in health care (Hunt et al., 2021; Kaambwa et al., 2012; Ward et al., 2020; Zhang et al., 2023), natural disasters (Jones et al., 2022), psychology (Gomer & Yuan, 2023), agriculture (Robbins et al., 2013), epidemiology (Allotey & Harel, 2019; Balzer et al., 2020; Tsiampalis & Panagiotakos, 2020), and crime (Blasco, 2021; Edwards et al., 2017; Stockton et al., 2024). A sensitivity analysis using the steps described in Figure 1 could be valuable across these and many other disciplines in which missing data convolutes the resulting analysis conclusions.

We recognize the inherent difficulty in assessing racial discrimination using observational studies with missing or poorly recorded data. We have provided a method for assessing the sensitivity of results that aim to measure racial disparities. In our own work we find that there is quite a bit of missing data in the SOPP that is not MCAR. We were able to find a substantial shift toward discrimination in considering the outcome test in Ohio. The sensitivity analysis on the ATE bounds show that there is anti-Black discrimination in multiple jurisdictions, even considering the missing race information.

## Acknowledgments

Amber Lee was important to much of the work, and we appreciate her early contributions. Additionally, we appreciate the ideas put forth by Ghassan Sarkis and the following students, who worked on aspects of the project: Oliver Chang, Emma Godfrey, Will Gray, Ethan Ong, Kyle Torres, Arm Wonghirundacha, and Ivy Yuan.

## Disclosure Statement

The authors have no conflicts of interest to declare. S.K. was supported in part by grants from the Pomona College SURP program, the Kenneth Cooke Summer Research Fellowship, and NSF Conference grant 2328509.

# Appendices

## Appendix A. Causal DAG

The directed acyclic graph (DAG) in Figure 3 illustrates missing data as well as missing potential outcomes mechanisms in the traffic stop data set. Here we provide detailed information about the nodes in the DAG. Additionally, Table A1 provides examples for causal relationships and missingness mechanisms that might exist in the DAG in Figure 3. We use $X$ to represent covariates. $X_1$ refers to completely observed covariates in the data set, such as date, time, location, county, and police department. These variables are complete, that is, 0% missingness, across a particular state patrol agency or municipal police department data set in the Stanford Open Policing Project (SOPP). $X_2$ comprises variables with missing data—subject race, subject age, stop outcome, and so on. The treatment, $M$, represents whether the individual was stopped by the police. As a mediating variable, $M$ may be affected by covariates $X$ (both complete and incomplete) such as date or subject race. The outcome, $Y$, represents any post-stop outcome such as whether the individual was searched, found carrying contraband, or arrested. Unmeasured confounders, $U$ (e.g., the officer's suspicion or their perception of the individual's race), might affect the treatment, $M$ (whether the individual is stopped), and the outcome, $Y$ (whether they are searched).

We introduce three indicator variables $R_{X_2}$, $R_M$, and $R_Y$ to denote whether an instance contains missing data for partially observed covariates $(X_2)$, the treatment $(M)$, or outcome $(Y)$. Moreno-Betancur et al. (2018) also include unmeasured causes of missingness $(W)$ that might affect the probability of missingness. $W$ could include factors such as the weather or departmental policy changes for the precinct that might affect data records.

Having no arrows to the missingness indicators, $M$, would mean the missingness mechanism is MCAR and entirely dictated by unmeasured factors, $W$. However, in the case of SOPP data, we notice that complete and missing covariates have the potential to affect the missingness of the stop, the outcome, and the covariates themselves as discussed in Section 3. For example, a certain police department might have higher rates of missing data in their reporting (Pierson et al., 2020) or officers might omit data at differing rates for different races (Chanin & Welsh, 2021). Hence, Figure 3 contains arrows from $X_1$ and $X_2$ to the three missingness indicators.

Additionally, policing data sets such as SOPP inherently only include observations with one level of the treatment—individuals who were stopped $(M = 1)$—and contain no report on the number of individuals who were observed but not stopped $(M = 0)$ (Knox et al., 2020).

Lastly, the stop outcome, $Y$, may affect its own missingness, $M_Y$, as well as records of covariates, $M_{Z_2}$, such as subject race (i.e., officers might omit race if the individual was not searched). These causal effects are depicted in the DAG.

### Table A1. Example causal relationships and missingness mechanisms to describe the directed acyclic graph in Figure 3.[7]

| | | Arrow to | | | | | | | | |
|---|---|---|---|---|---|---|---|---|---|---|
| | | $X_1$ | $X_2$ | $M$ | $Y$ | $U$ | $R_{X_2}$ | $R_M$ | $R_Y$ | $W$ |
| **Arrow from** | $X_!$ | | | More stops in some locations | More searches in some locations | | More race missingness early morning | Less stops early morning | Less searches early morning | |
| | $X_2$ | | | More stops among Black civilians | More searches among Black civilians | | More race missingness for Black stops (Chanin & Welsh, 2021) | Less stops among White civilians | Complacency in search records for some race | |
| | $M$ | | | | If someone is not stopped, they cannot be searched | | If someone is not stopped, their race records are missing | If someone is not stopped, their stop records are missing | | |
| | $Y$ | | | | | | Complacency in race records if not searched | Complacency in stop records if not searched | Complacency in search records if not searched | |
| | $U$ | | | Police suspicion | Police suspicion | | | | | |
| | $R_{X_2}$ | | | | | | | | | |
| | $R_M$ | | | | | | | | | |
| | $R_Y$ | | | | | | | | | |
| | $W$ | | | | | | | | | |

# Appendix B. Stanford Open Policing Project Data Summary

**Table B1. The 20 most commonly recorded variables in the Stanford Open Policing Project.**

| Covariate | Number of data sets |
|---|---|
| date | 88 |
| subject race | 77 |
| outcome | 74 |
| location | 73 |
| time | 72 |
| subject sex | 71 |
| citation issued | 68 |
| subject age | 53 |
| latitude | 51 |
| longitude | 51 |
| warning issued | 49 |
| arrest made | 48 |
| search conducted | 45 |
| violation | 45 |
| officer id hash | 43 |
| contraband found | 36 |
| search basis | 34 |
| reason for stop | 33 |
| county name | 31 |
| vehicle make | 29 |

**Table B2. Stanford Open Policing Project (SOPP) per variable missingness rates: List of the 88 data sets in the SOPP. Eight bold data sets were selected based on the criteria discussed in Section 3.4. `NA` indicates that the variable was not recorded.**

| Data set | Date_range | #Rows | #Covariates | subject_race | search_ conducted | contraband_found |
|---|---|---|---|---|---|---|
| AR, Little Rock | 2017–2017 | 13,641 | 13 | 0.051 | NA | NA |
| AZ, Gilbert | 2008–2018 | 480,599 | 12 | NA | NA | NA |
| AZ, Mesa | 2014–2019 | 157,015 | 20 | 0.000 | NA | NA |
| AZ, Statewide | 2009–2017 | 3,498,159 | 29 | 0.116 | 0.000 | 0.000 |
| CA, Anaheim | 2012–2017 | 87,876 | 4 | NA | NA | NA |
| CA, Bakersfield | 2008–2018 | 189,685 | 18 | 0.441 | NA | NA |
| CA, Long Beach | 2008–2017 | 365,924 | 27 | 0.000 | NA | NA |
| CA, Los Angeles | 2010–2018 | 5,418,402 | 10 | 0.000 | NA | NA |
| CA, Oakland | 2013–2017 | 133,407 | 28 | 0.000 | 0.000 | 0.000 |
| CA, San Bernardino | 2011–2017 | 90,523 | 12 | NA | NA | NA |
| CA, San Diego | 2014–2017 | 383,027 | 21 | 0.322 | 0.000 | 0.000 |
| CA, San Francisco | 2007–2016 | 905,070 | 22 | 0.000 | 0.000 | 0.000 |
| CA, San Jose | 2013–2018 | 152,834 | 20 | 1.782 | 1.618 | 0.000 |
| CA, Santa Ana | 2014–2018 | 46,268 | 15 | 0.225 | NA | NA |
| **CA, Statewide** | **2009–2016** | **31,778,515** | **22** | **0.000** | **0.000** | **95.692** |
| CA, Stockton | 2012–2016 | 41,629 | 17 | 0.495 | 0.000 | NA |
| CO, Aurora | 2012–2020 | 257,606 | 16 | 0.146 | NA | NA |
| CO, Denver | 2010–2018 | 1,870,731 | 14 | NA | NA | NA |
| **CO, Statewide** | **2010–2017** | **3,112,853** | **20** | **12.903** | **7.424** | **0.000** |
| CT, Hartford | 2013–2016 | 18,439 | 26 | 0.000 | 0.000 | 0.058 |
| CT, Statewide | 2013–2015 | 1,175,339 | 26 | 0.000 | 0.000 | 0.000 |
| FL, Saint Petersburg | 2010–2010 | 16,383 | 9 | NA | NA | NA |
| FL, Statewide | 2010–2018 | 7,297,538 | 34 | 0.025 | 31.516 | NA |
| FL, Tampa | 1973–2018 | 2,818,240 | 12 | 0.000 | NA | NA |
| GA, Statewide | 2012–2016 | 1,906,772 | 19 | 47.112 | NA | NA |
| IA, Statewide | 2006–2016 | 2,441,335 | 20 | 69.527 | NA | NA |
| ID, Idaho Falls | 2008–2016 | 104,814 | 13 | NA | NA | NA |
| **IL, Chicago** | **2012–2020** | **2,108,098** | **35** | **31.343** | **0.000** | **0.000** |
| IL, Statewide | 2012–2017 | 12,748,173 | 29 | 0.006 | 0.132 | 0.802 |
| IN, Fort Wayne | 2007–2017 | 265,019 | 12 | NA | NA | NA |
| KS, Wichita | 2006–2020 | 1,030,376 | 22 | 6.535 | NA | NA |
| KY, Louisville | 2015–2020 | 146,562 | 33 | 0.058 | 0.000 | NA |
| KY, Owensboro | 2015–2017 | 6,921 | 18 | 0.257 | NA | NA |
| LA, New Orleans | 2010–2018 | 512,092 | 32 | 3.339 | 0.000 | 0.000 |
| MA, Statewide | 2007–2015 | 3,416,238 | 24 | 0.049 | 0.000 | 0.000 |
| MD, Baltimore | 2011–2017 | 854,759 | 9 | NA | NA | NA |
| **MD, Statewide** | **2007–2014** | **3,669,665** | **28** | **0.380** | **0.000** | **17.976** |
| MI, Statewide | 2001–2016 | 800,302 | 20 | 2.185 | NA | NA |
| MN, Saint Paul | 2001–2016 | 675,156 | 16 | 17.477 | 0.000 | NA |
| MO, Statewide | 2010–2015 | 10,038,706 | 9 | 0.000 | 0.000 | 0.000 |
| MS, Statewide | 2013–2016 | 758,412 | 13 | 0.055 | NA | NA |
| MT, Statewide | 2009–2017 | 921,228 | 30 | 0.012 | 0.000 | NA |
| NC, Charlotte | 2000–2015 | 1,598,453 | 29 | 0.000 | 0.000 | 0.000 |
| NC, Durham | 2001–2015 | 326,024 | 29 | 0.000 | 0.000 | 0.000 |
| NC, Fayetteville | 2000–2015 | 486,998 | 29 | 0.000 | 0.000 | 0.000 |
| NC, Greensboro | 2000–2015 | 600,031 | 29 | 0.001 | 0.000 | 0.000 |
| NC, Raleigh | 2002–2015 | 856,400 | 29 | 0.000 | 0.000 | 0.000 |
| NC, Statewide | 2000–2015 | 20,286,645 | 29 | 0.000 | 0.000 | 0.000 |
| NC, Winston-Salem | 2007–2016 | 452,560 | 29 | 0.000 | 0.000 | 0.000 |
| ND, Grand Forks | 2010–2015 | 82,535 | 14 | 1.645 | NA | NA |
| ND, Statewide | 2010–2015 | 330,132 | 12 | 0.037 | NA | NA |
| NE, Statewide | 2002–2016 | 9,031,494 | 10 | 0.000 | 0.000 | NA |
| NH, Statewide | 2014–2015 | 259,822 | 17 | 36.077 | NA | NA |
| NJ, Camden | 2013–2020 | 195,298 | 23 | 0.838 | NA | NA |
| **NJ, Statewide** | **2009–2016** | **3,845,334** | **24** | **96.142** | **96.306** | **4.435** |
| NV, Henderson | 2011–2019 | 111,090 | 19 | 2.574 | NA | NA |
| NV, Statewide | 2012–2016 | 737,285 | 11 | 0.068 | NA | NA |
| NY, Albany | 2008–2017 | 74,093 | 16 | 32.251 | NA | NA |
| NY, Statewide | 2010–2017 | 7,962,169 | 19 | 0.000 | NA | NA |
| OH, Cincinnati | 2009–2018 | 315,281 | 25 | 0.000 | NA | NA |
| OH, Columbus | 2012–2016 | 128,157 | 18 | 0.000 | 0.000 | NA |
| **OH, Statewide** | **2010–2017** | **7,225,577** | **24** | **8.704** | **0.000** | **80.163** |
| OK, Oklahoma City | 2011–2020 | 945,107 | 27 | 0.281 | NA | NA |
| OK, Tulsa | 2009–2016 | 438,248 | 19 | 0.974 | NA | NA |
| OR, Statewide | 2010–2014 | 1,143,017 | 5 | 0.000 | NA | NA |
| PA, Philadelphia | 2014–2018 | 1,865,096 | 22 | 0.000 | 0.000 | 0.000 |
| RI, Statewide | 2005–2015 | 509,681 | 31 | 5.704 | 0.000 | 0.000 |
| SC, Statewide | 2005–2016 | 8,983,810 | 34 | 0.000 | 0.000 | 0.000 |
| SD, Statewide | 2012–2016 | 435,895 | 16 | NA | NA | NA |
| TN, Nashville | 2010–2019 | 3,092,351 | 42 | 0.060 | 0.001 | 0.000 |
| TN, Statewide | 1971–2016 | 3,829,082 | 20 | 0.798 | NA | NA |
| TX, Arlington | 2016–2016 | 112,526 | 19 | 0.000 | 0.000 | NA |
| TX, Austin | 2006–2016 | 483,255 | 29 | 0.000 | 0.000 | 0.000 |
| TX, Garland | 2012–2019 | 159,845 | 23 | 0.014 | NA | NA |
| TX, Houston | 2014–2020 | 2,045,972 | 21 | 20.168 | NA | NA |
| TX, Lubbock | 2008–2018 | 546,101 | 10 | NA | NA | NA |
| TX, Plano | 2012–2015 | 249,043 | 36 | 0.000 | 0.049 | 0.000 |
| TX, San Antonio | 2012–2020 | 1,301,103 | 33 | 0.037 | 0.000 | 0.000 |
| TX, Statewide | 2006–2017 | 27,426,840 | 34 | 0.001 | 8.018 | 0.000 |
| VA, Statewide | 2006–2016 | 5,006,847 | 11 | 0.000 | 0.000 | NA |
| VT, Burlington | 2012–2020 | 36,845 | 26 | 4.627 | 0.000 | 0.000 |
| VT, Statewide | 2010–2015 | 283,285 | 25 | 1.406 | 0.000 | 0.000 |
| WA, Seattle | 2006–2015 | 319,959 | 27 | 99.896 | NA | NA |
| **WA, Statewide** | **2009–2018** | **11,333,425** | **30** | **28.176** | **0.000** | **0.000** |
| WA, Tacoma | 2007–2017 | 271,912 | 15 | NA | NA | NA |
| WI, Madison | 2007–2020 | 331,450 | 25 | 1.122 | NA | NA |
| **WI, Statewide** | **2010–2016** | **1,058,902** | **44** | **14.505** | **0.000** | **0.000** |
| WY, Statewide | 2011–2012 | 173,311 | 16 | 0.294 | NA | NA |

# Appendix C. Outcome Test: Additional Results

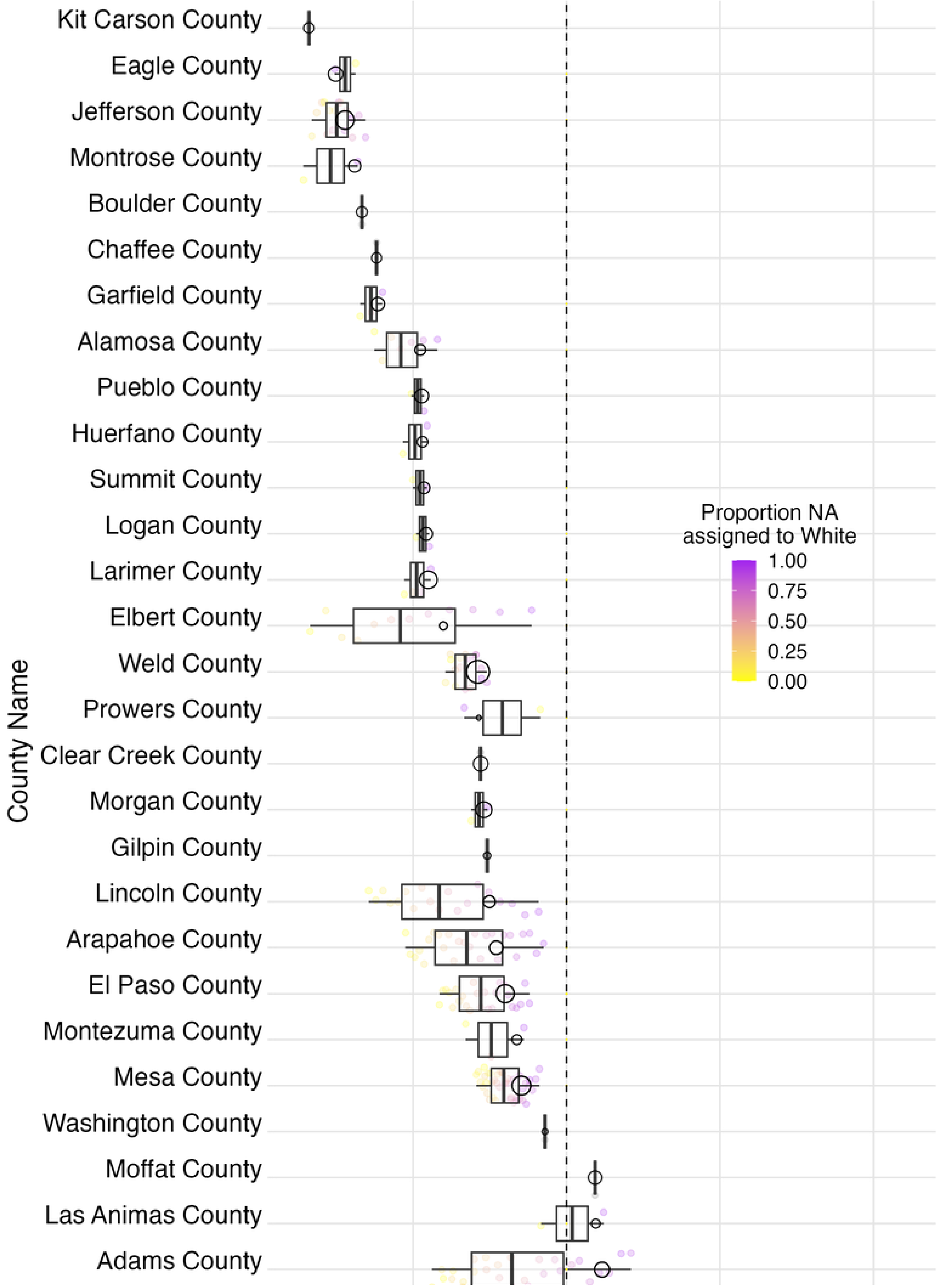

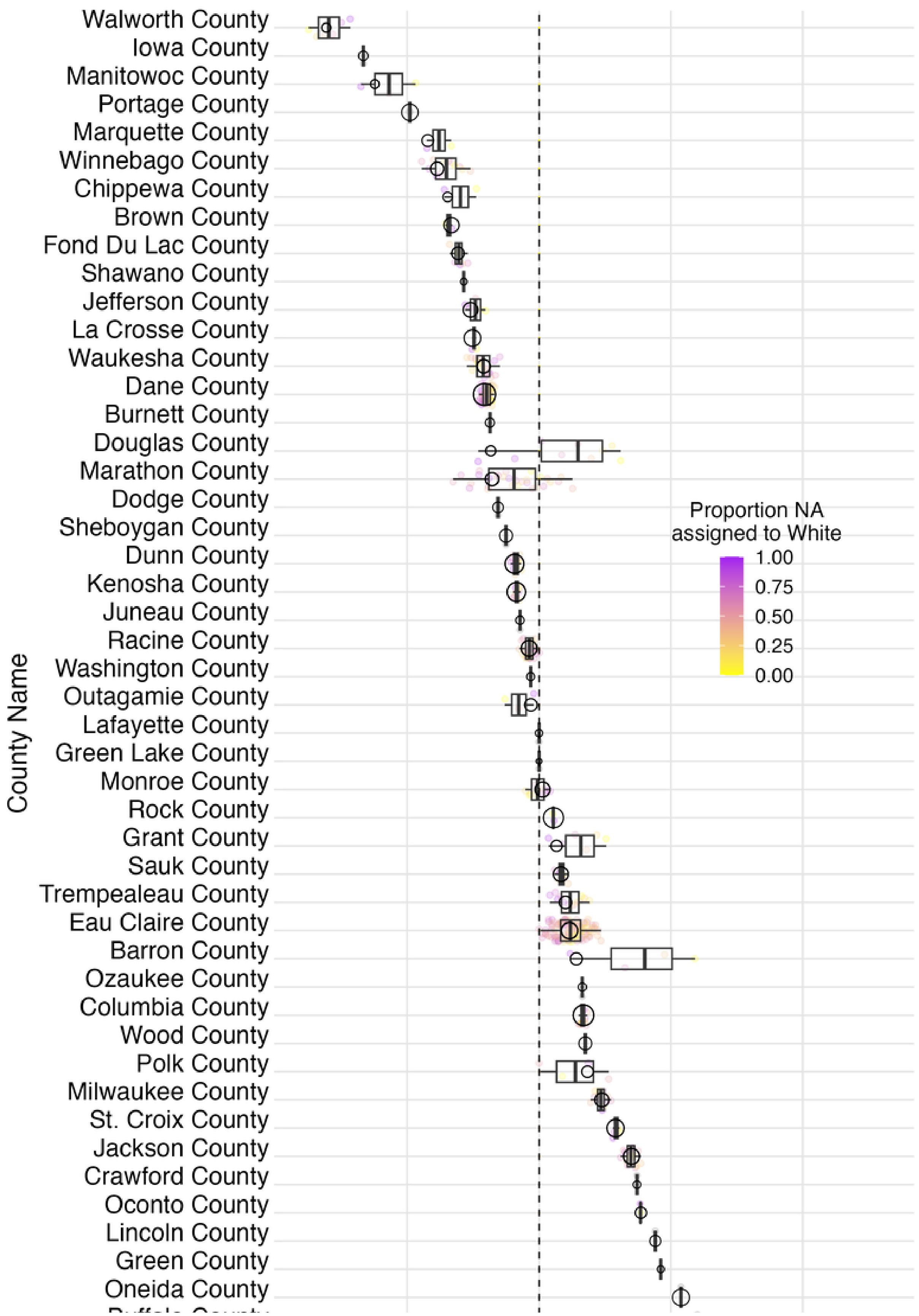
County Name
Walworth County
Iowa County
Manitowoc County
Portage County
Marquette County
Winnebago County
Chippewa County
Brown County
Fond Du Lac County
Shawano County
Jefferson County
La Crosse County
Waukesha County
Dane County
Burnett County
Douglas County
Marathon County
Dodge County
Sheboygan County
Dunn County
Kenosha County
Juneau County
Racine County
Washington County
Outagamie County
Lafayette County
Green Lake County
Monroe County
Rock County
Grant County
Sauk County
Trempealeau County
Eau Claire County
Barron County
Ozaukee County
Columbia County
Wood County
Polk County
Milwaukee County
St. Croix County
Jackson County
Crawford County
Oconto County
Lincoln County
Green County
Oneida County
Proportion NA assigned to White
1.00
0.75
0.50
0.25
0.00

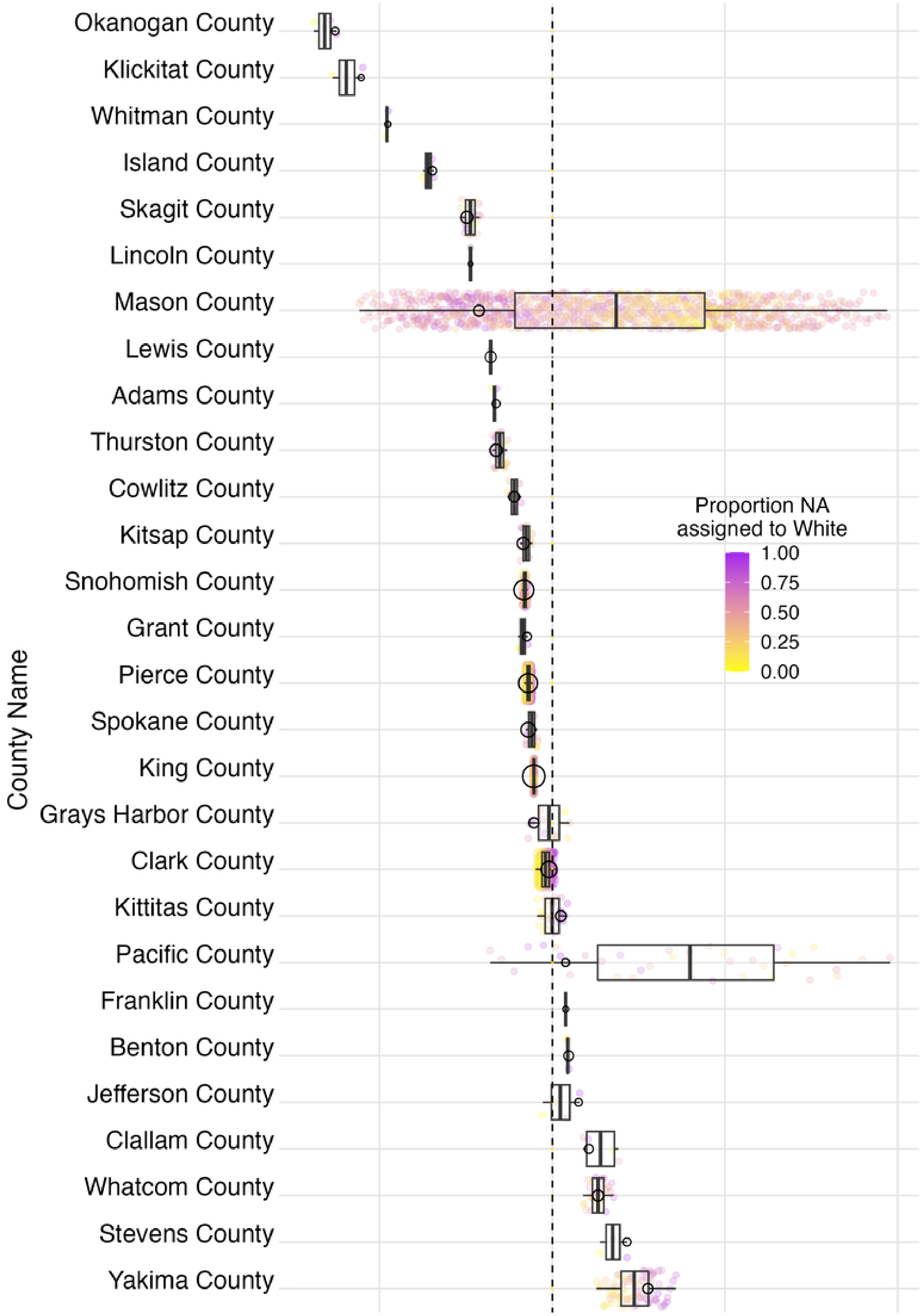
County Name
Okanogan County
Klickitat County
Whitman County
Island County
Skagit County
Lincoln County
Mason County
Lewis County
Adams County
Thurston County
Cowlitz County
Kitsap County
Snohomish County
Grant County
Pierce County
Spokane County
King County
Grays Harbor County
Clark County
Kittitas County
Pacific County
Franklin County
Benton County
Jefferson County
Clallam County
Whatcom County
Stevens County
Yakima County
Proportion NA assigned to White
1.00
0.75
0.50
0.25
0.00

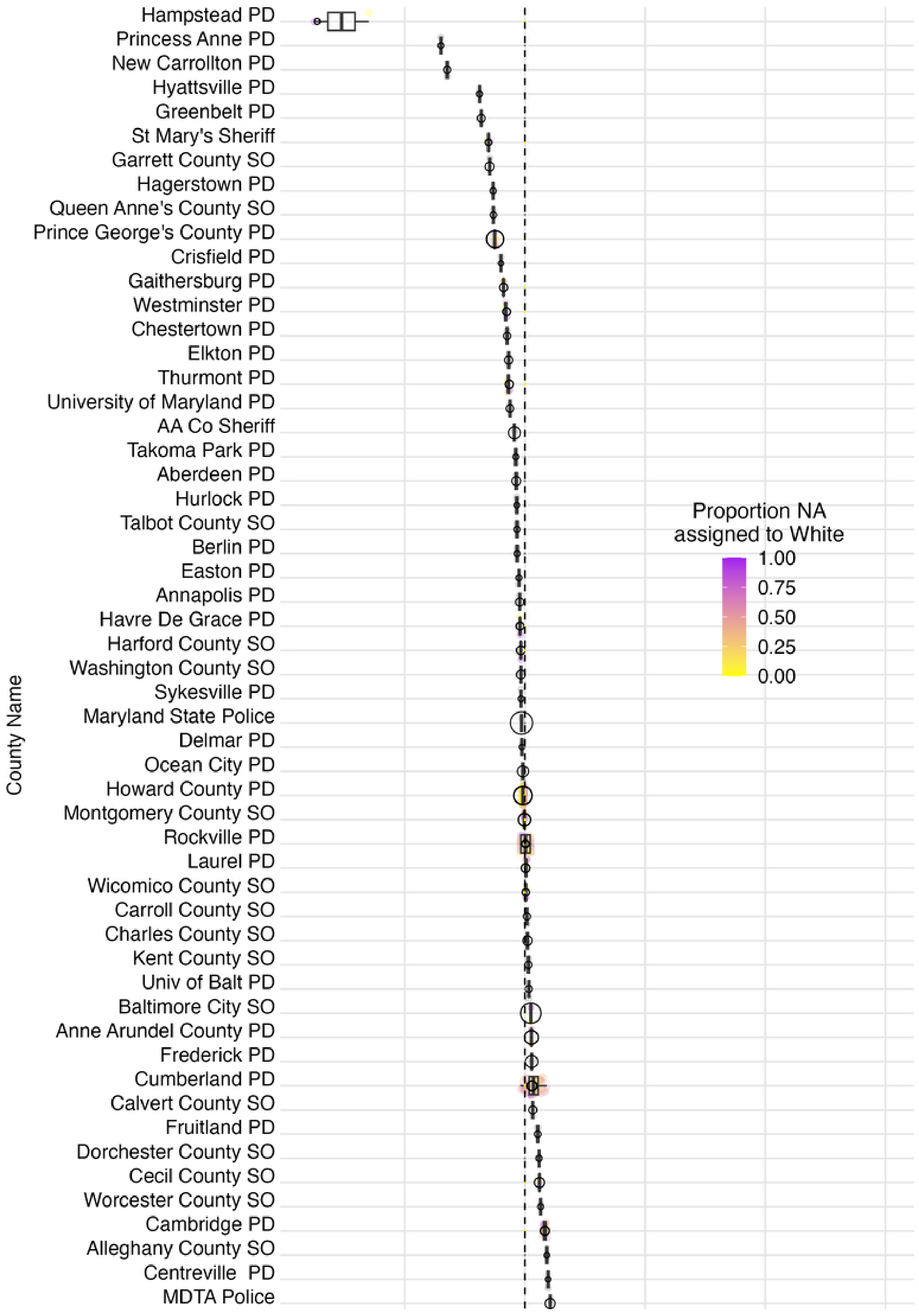
County Name
Hampstead PD
Princess Anne PD
New Carrollton PD
Hyattsville PD
Greenbelt PD
St Mary's Sheriff
Garrett County SO
Hagerstown PD
Queen Anne's County SO
Prince George's County PD
Crisfield PD
Gaithersburg PD
Westminster PD
Chestertown PD
Elkton PD
Thurmont PD
University of Maryland PD
AA Co Sheriff
Takoma Park PD
Aberdeen PD
Hurlock PD
Talbot County SO
Berlin PD
Easton PD
Annapolis PD
Havre De Grace PD
Harford County SO
Washington County SO
Sykesville PD
Maryland State Police
Delmar PD
Ocean City PD
Howard County PD
Montgomery County SO
Rockville PD
Laurel PD
Wicomico County SO
Carroll County SO
Charles County SO
Kent County SO
Univ of Balt PD
Baltimore City SO
Anne Arundel County PD
Frederick PD
Cumberland PD
Calvert County SO
Fruitland PD
Dorchester County SO
Cecil County SO
Worcester County SO
Cambridge PD
Alleghany County SO
Centreville PD
MDTA Police
Proportion NA assigned to White
1.00
0.75
0.50
0.25
0.00

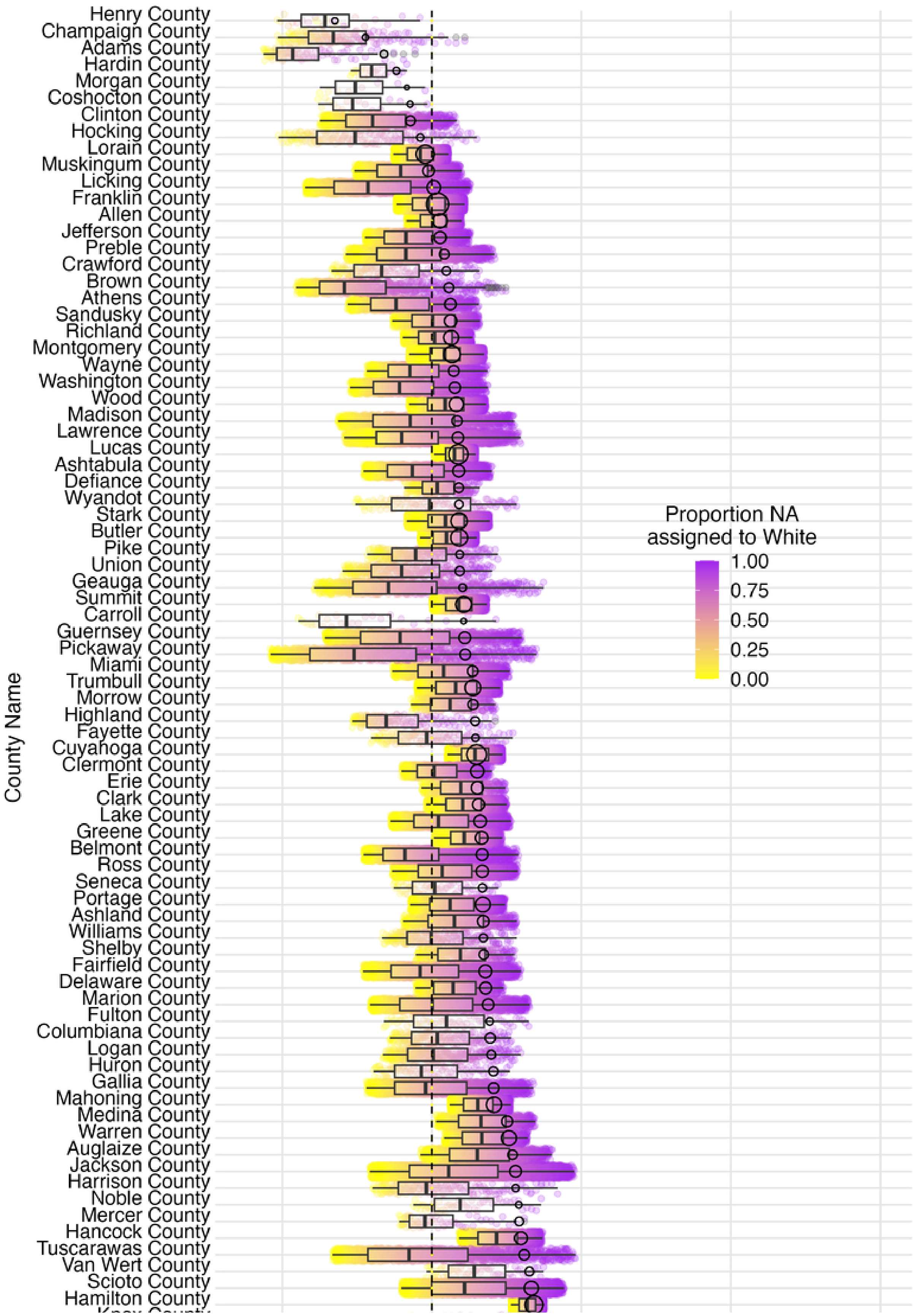
County Name
Henry County
Champaign County
Adams County
Hardin County
Morgan County
Coshocton County
Clinton County
Hocking County
Lorain County
Muskingum County
Licking County
Franklin County
Allen County
Jefferson County
Preble County
Crawford County
Brown County
Athens County
Sandusky County
Richland County
Montgomery County
Wayne County
Washington County
Wood County
Madison County
Lawrence County
Lucas County
Ashtabula County
Defiance County
Wyandot County
Stark County
Butler County
Pike County
Union County
Geauga County
Summit County
Carroll County
Guernsey County
Pickaway County
Miami County
Trumbull County
Morrow County
Highland County
Fayette County
Cuyahoga County
Clermont County
Erie County
Clark County
Lake County
Greene County
Belmont County
Ross County
Seneca County
Portage County
Ashland County
Williams County
Shelby County
Fairfield County
Delaware County
Marion County
Fulton County
Columbiana County
Logan County
Huron County
Gallia County
Mahoning County
Medina County
Warren County
Auglaize County
Jackson County
Harrison County
Noble County
Mercer County
Hancock County
Tuscarawas County
Van Wert County
Scioto County
Hamilton County
Proportion NA
assigned to White
1.00
0.75
0.50
0.25
0.00

## Appendix D. Average Treatment Effect Bounds: Additional Results

### D.1. Random Assignment

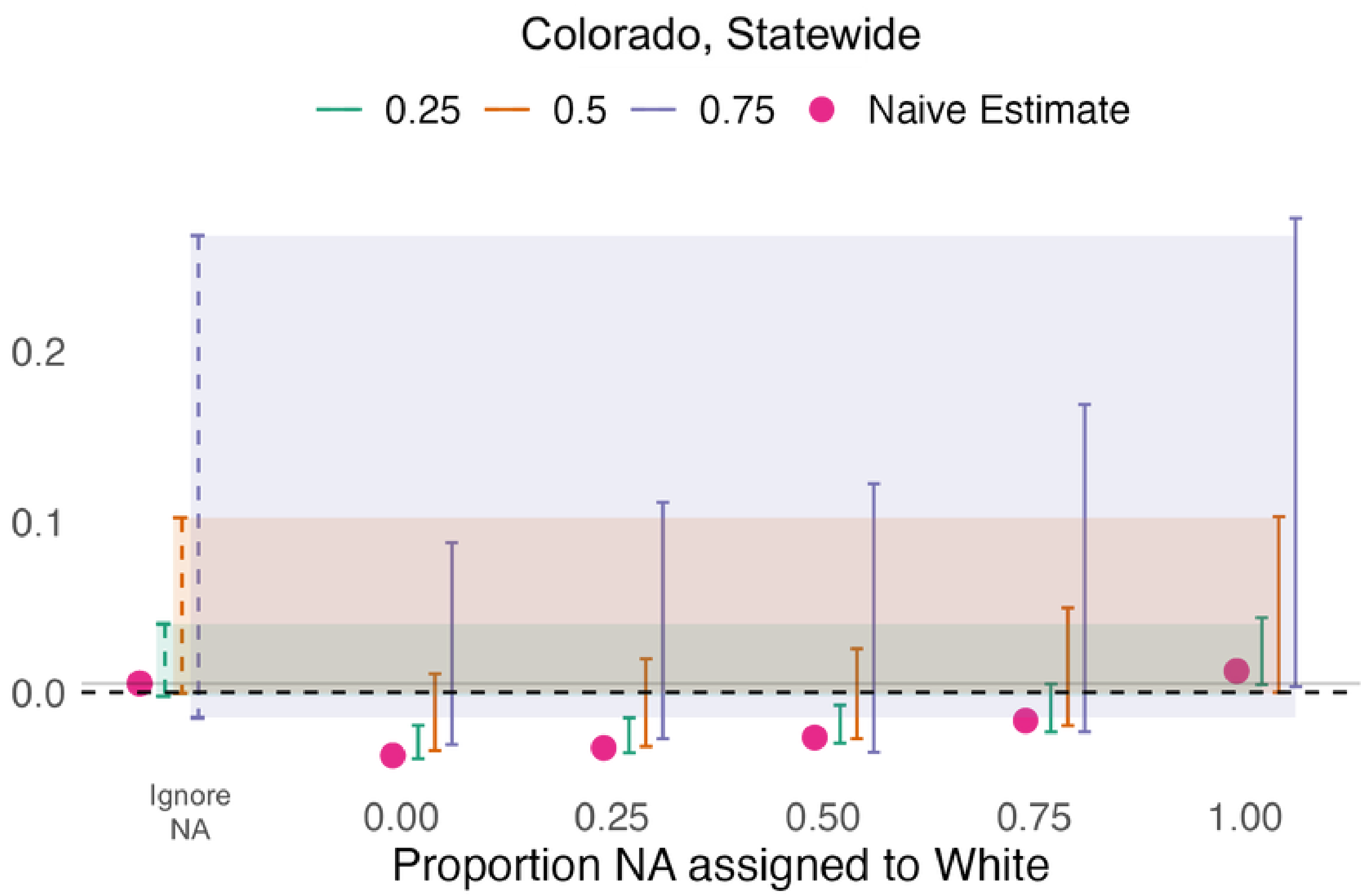

**Figure D1.** Dashed line at zero indicates no differential racial impact. Disparity is the difference in proportion of search for Black minus proportion of search for White. Disparity is measured as bounds on the search average treatment effect, given recorded outcome, for different levels of the ρ parameter (the overall proportion of racially discriminatory stops). The *x*-axis of each plot presents a different allocation of assigning `NA` race observations to either Black or White. `NA` values are missing observations.

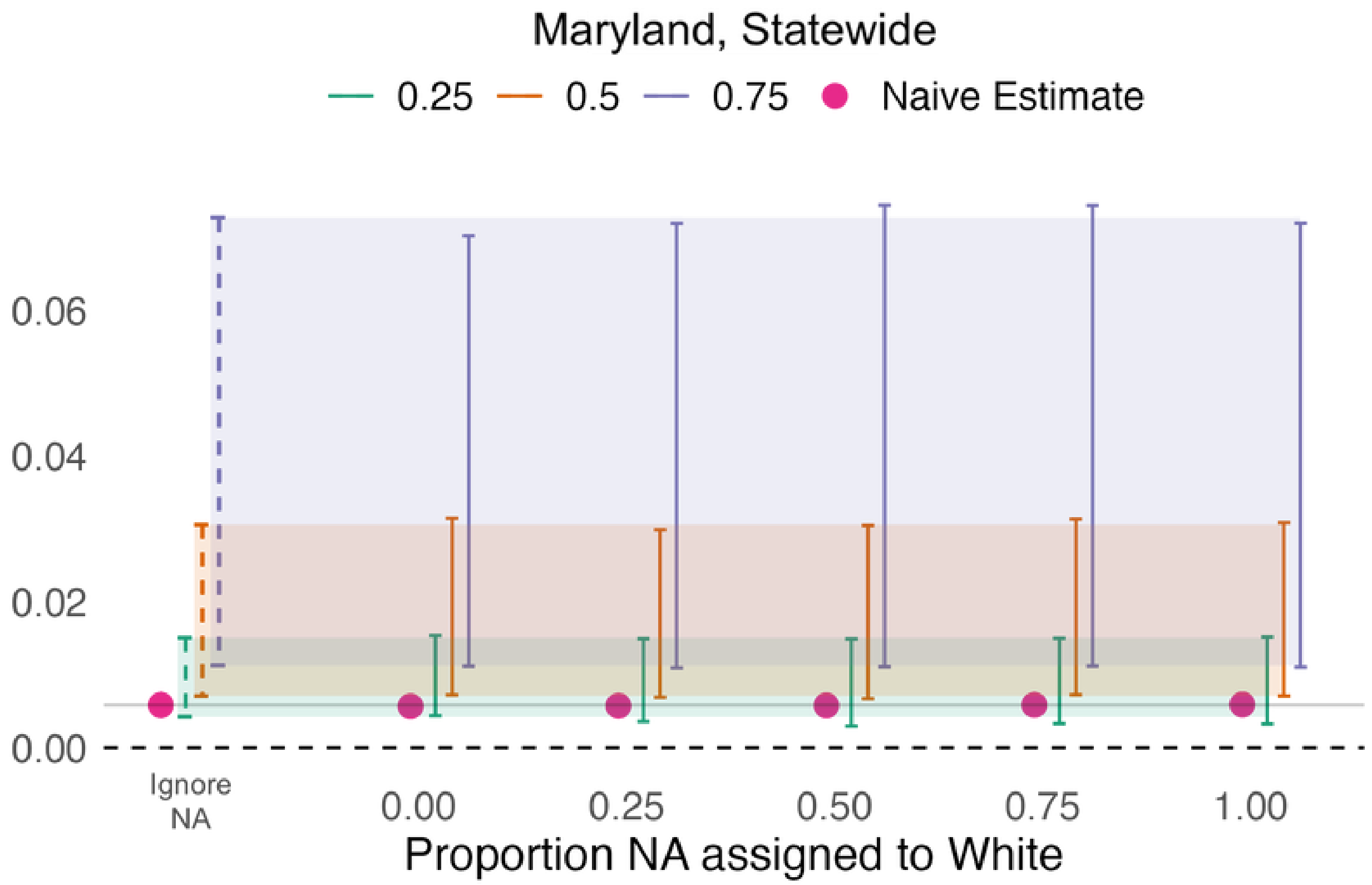

**Figure D2.** Dashed line at zero indicates no differential racial impact. Disparity is the difference in proportion of search for Black minus proportion of search for White. Disparity is measured as bounds on the search average treatment effect, given recorded outcome, for different levels of the ρ parameter (the overall proportion of racially discriminatory stops). The *x*-axis of each plot presents a different allocation of assigning `NA` race observations to either Black or White. `NA` values are missing observations.

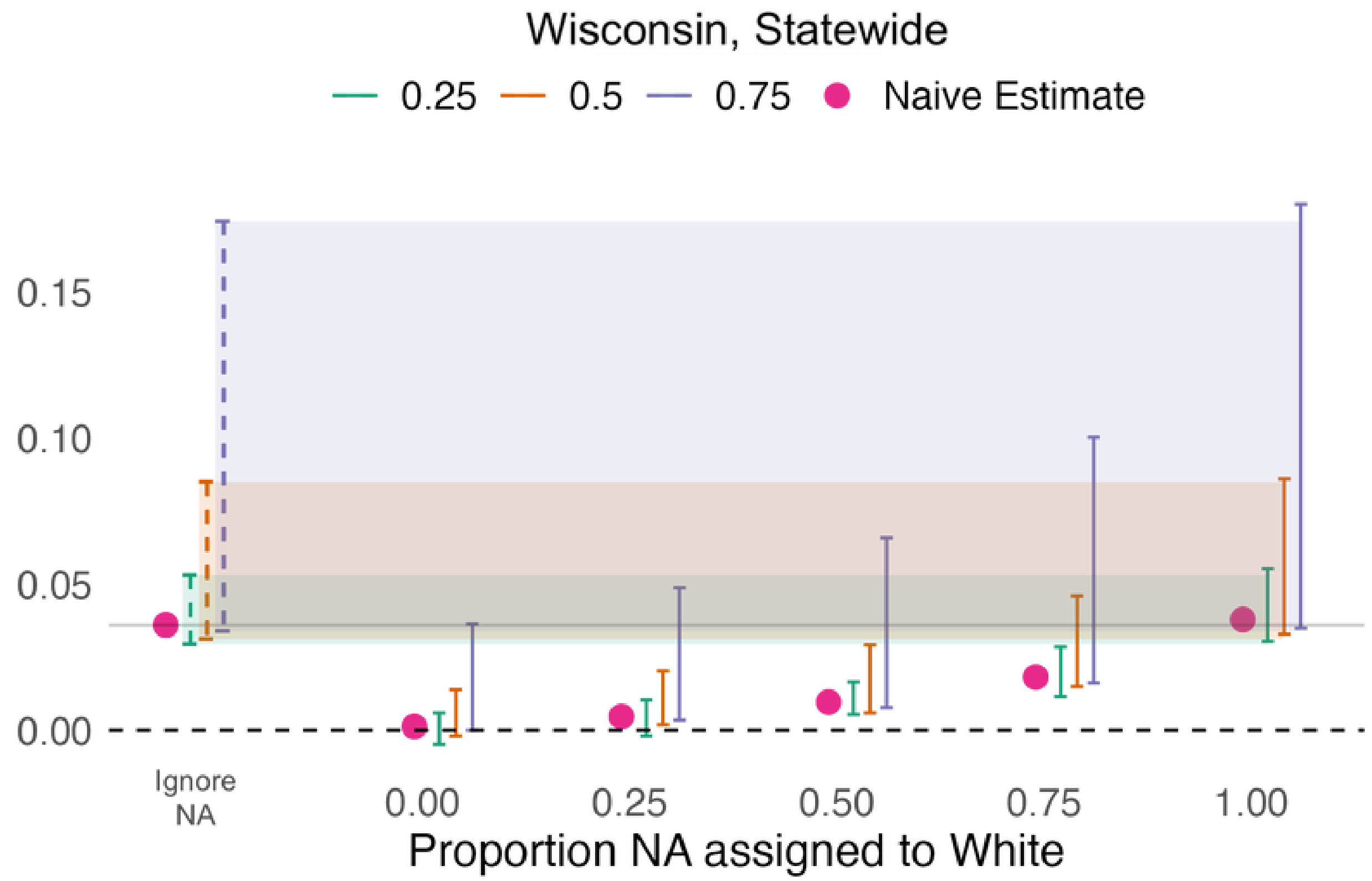

**Figure D3.** Dashed line at zero indicates no differential racial impact. Disparity is the difference in proportion of search for Black minus proportion of search for White. Disparity is measured as bounds on the search average treatment effect, given recorded outcome, for different levels of the ρ parameter (the overall proportion of racially discriminatory stops). The *x*-axis of each plot presents a different allocation of assigning `NA` race observations to either Black or White. `NA` values are missing observations.

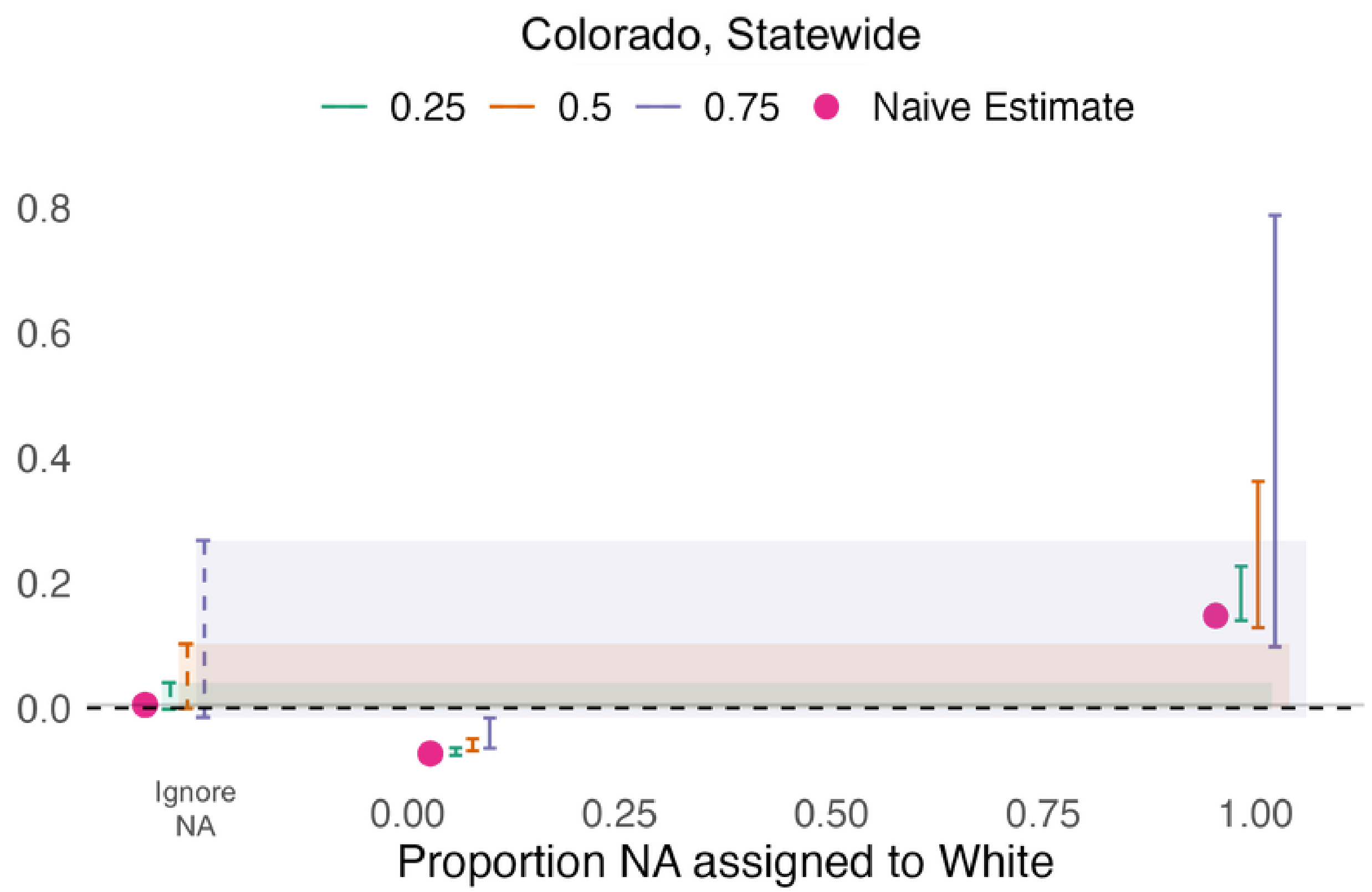

**Figure D4.** Dashed line at zero indicates no differential racial impact. Disparity is the difference in proportion of search for Black minus proportion of search for White. Disparity is measured as bounds on the search average treatment effect, given recorded outcome, for different levels of the ρ parameter (the overall proportion of racially discriminatory stops). The *x*-axis of each plot presents the *ignore NA* case as well as the two most extreme scenarios of allocating `NA` race observations to either Black or White. Note that for Colorado, statewide, the disparity estimates can change direction from discrimination against Black drivers to the opposite direction. `NA` values are missing observations.

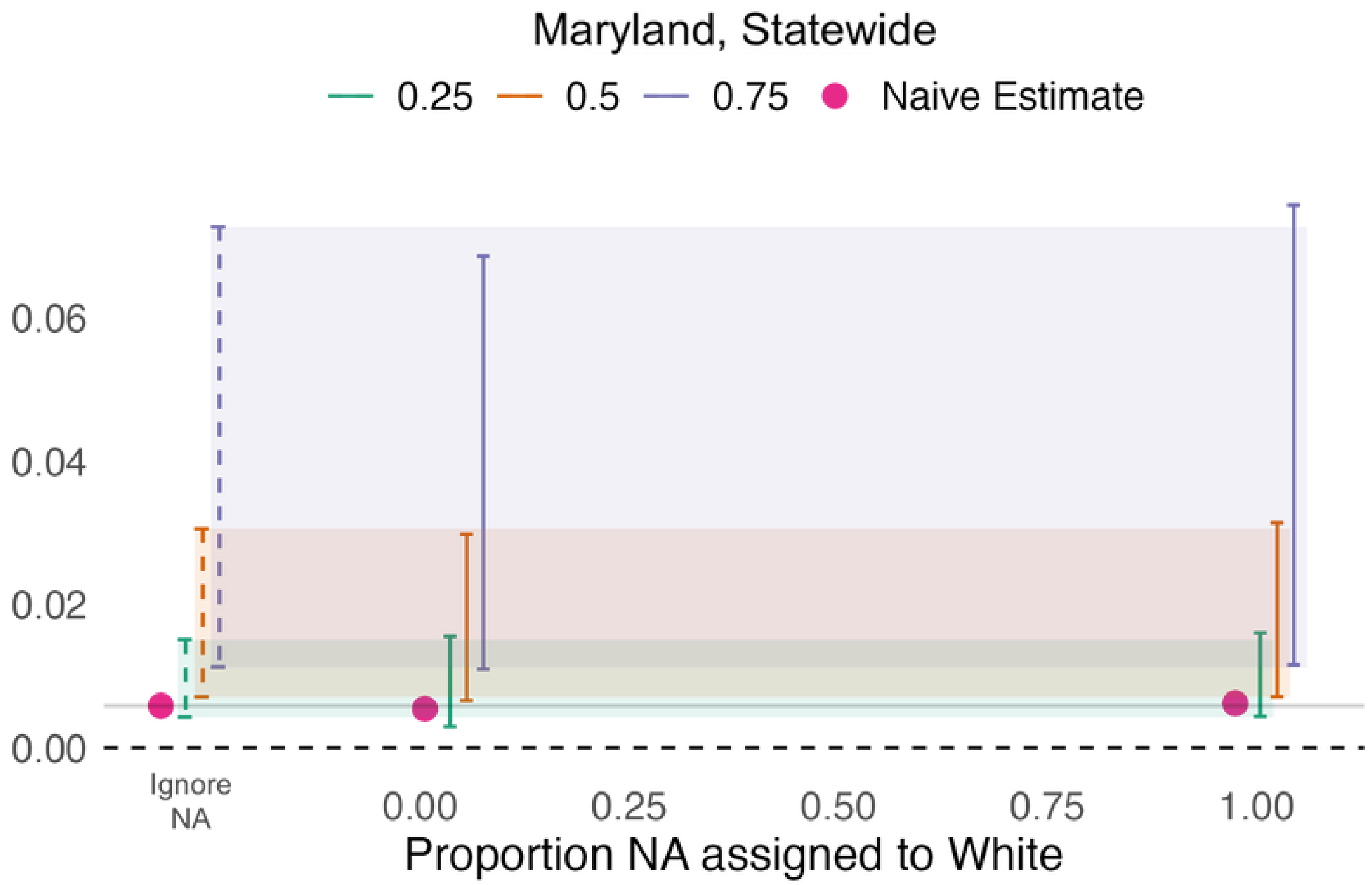

**Figure D5.** Dashed line at zero indicates no differential racial impact. Disparity is the difference in proportion of search for Black minus proportion of search for White. Disparity is measured as bounds on the search average treatment effect (ATE), given recorded outcome, for different levels of the ρ parameter (the overall proportion of racially discriminatory stops). The *x*-axis of each plot presents the *ignore NA* case as well as the two most extreme scenarios of allocating `NA` race observations to either Black or White. Note that for Maryland, statewide, the ATE bounds are always above zero, even in the most extreme cases, indicating racial discrimination against Black drivers. `NA` values are missing observations.

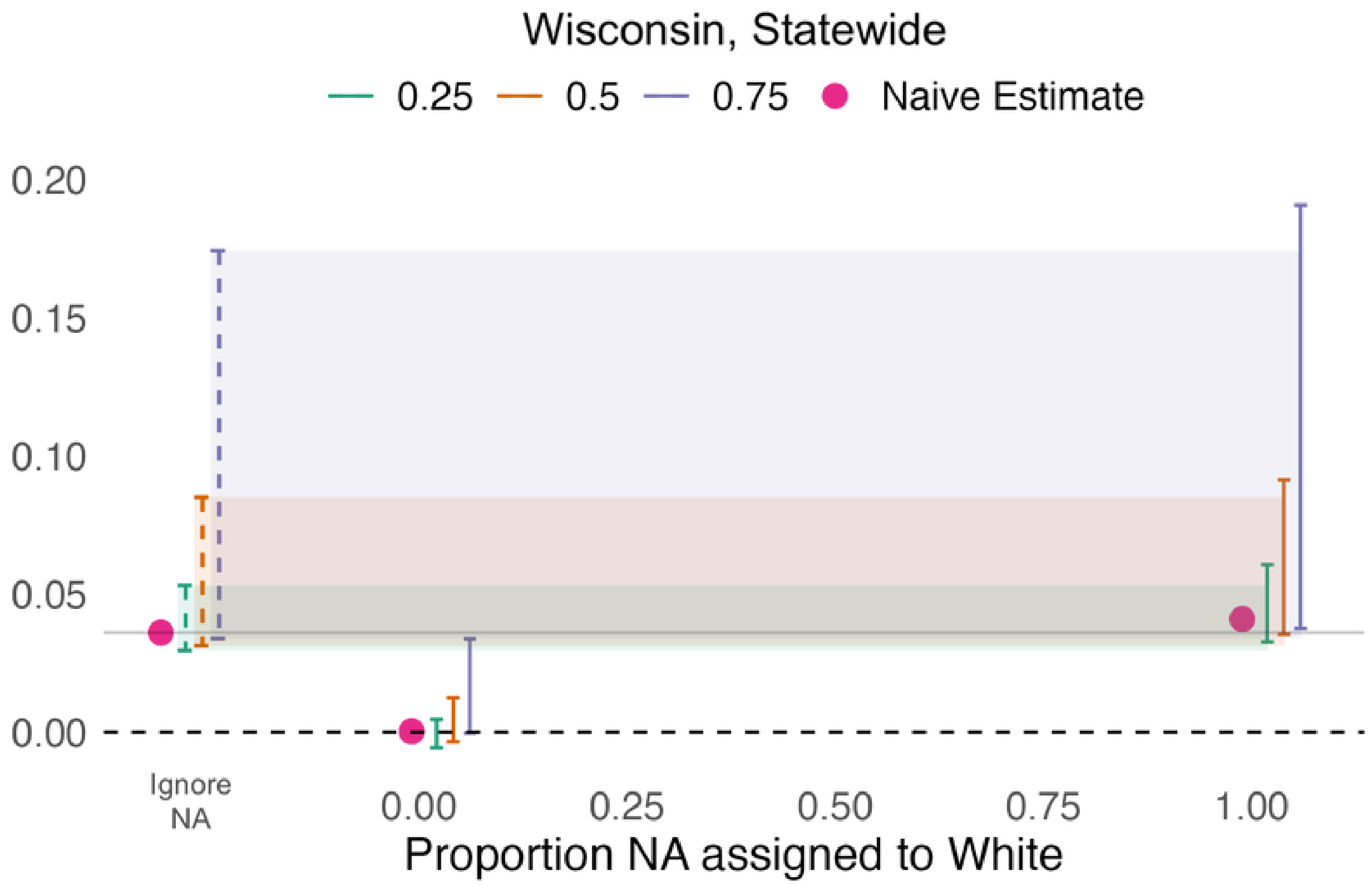

**Figure D6.** Dashed line at zero indicates no differential racial impact. Disparity is the difference in proportion of search for Black minus proportion of search for White. Disparity is measured as bounds on the search average treatment effect (ATE), given recorded outcome, for different levels of the ρ parameter (the overall proportion of racially discriminatory stops). The *x*-axis of each plot presents the *ignore NA* case as well as the two most extreme scenarios of allocating `NA` race observations to either Black or White. Note that for Wisconsin, statewide, the ATE bounds are almost always above zero, indicating discrimination against Black drivers. `NA` values are missing observations.

---

## Data Repository/Code

The Open Policing Project data sets can be found at https://openpolicing.stanford.edu/data/. Code to replicate our findings is available at https://github.com/saatvikkher/missing-traffic

---

## Footnotes

1. https://openpolicing.stanford.edu/ ↩

2. After our manuscript was submitted, SOPP substantially expanded their database, in July 2025, and included a Python library for accessing the data: https://openpolicedata.readthedocs.io/. ↩

3. There are 88 data sets, so perhaps one could write $X^{(d)}$ for $1 \le d \le 88$. But the definitions only apply to one data set at a time, so we avoid the extra notation. ↩

4. The $m - 1$ in the denominator is because the missingness of the conditioning variable itself is not considered. ↩

5. Table scrolls horizontally for additional columns. ↩

6. Table scrolls horizontally for additional columns. ↩

7. Table scrolls horizontally for additional columns. ↩